\documentclass[prd,twocolumn,floatfix,amsmath,nofootinbib,amssymb,floatfix]{revtex4}
\usepackage{graphicx,color,dcolumn,booktabs,bm}
\usepackage{longtable,lscape}
\usepackage{pdfpages}
\usepackage{txfonts}
\usepackage{overpic}
\usepackage{amssymb}
\usepackage{makecell}
\usepackage{indentfirst}
\usepackage{feynmf}   
\usepackage{slashed}  
\usepackage{cases}
\usepackage{color}
\usepackage{multirow}
\usepackage{threeparttable}
\usepackage{enumerate}
\usepackage{subfigure}
\usepackage{diagbox}
\usepackage{mathrsfs}
\usepackage{cancel}
\usepackage{float}
\usepackage[colorlinks,
            citecolor=blue,
            anchorcolor=red,
            menucolor=red,
            linkcolor=red,
            filecolor=red,
            runcolor=red,
            urlcolor=blue,
            frenchlinks=red]{hyperref}

\usepackage{epstopdf}
\usepackage{amsmath}
\usepackage{dutchcal}
\usepackage{array}
\usepackage{booktabs}
\usepackage{makecell}
\usepackage{placeins}
\usepackage{textcomp}
\usepackage{amsmath}
\DeclareRobustCommand{\perthousand}{%
	\ifmmode
	\text{\textperthousand}%
	\else
	\textperthousand
	\fi}
\begin{document}
\title{Systematic analysis of the form factors of $B_c\rightarrow\eta_c$, $J/\psi$ and corresponding weak decays}
\author{Bin Wu$^{1}$}
\author{Guo-Liang Yu$^{1,2}$}
\email{yuguoliang2011@163.com}
\author{Jie Lu$^{1,2}$}
\author{Zhi-Gang Wang$^{1}$}
\email{zgwang@aliyun.com}

\affiliation{$^1$ Department of Mathematics and Physics, North China
Electric Power University, Baoding 071003, People's Republic of
China\\$^2$ Hebei Key Laboratory of Physics and Energy Technology, North China Electric Power University, Baoding 071000, China}
\date{\today}

\begin{abstract}
 The form factors of $B_c\rightarrow\eta_c$ and $B_c\rightarrow J/\psi$ are analyzed in the framework of three-point QCD sum rules. In these analyses, the contributions of the vacuum condensate terms $\langle g_{s}^{2}GG\rangle$ and $\langle g_{s}^{3}GGGf\rangle$ are considered. In addition, the decay widths and branching ratios of several decay channels are obtained by using the calculated form factors. These decay processes include the nonleptonic decays of $B_c^- \to \eta_c \pi^-$, $\eta_c K^-$, $\eta_c \rho^-$, $\eta_c K^{*-}$, $B_c^- \to J/\psi \pi^-$, $J/\psi K^-$, $J/\psi \rho^-$, $J/\psi K^{*-}$, and the semileptonic decays of $B_c^- \to \eta_c \mathcal{l} \bar{\nu}$, $B_c^- \to J/\psi \mathcal{l} \bar{\nu}$. These results about the form factors and decay properties of $B_c$ meson provide useful information for us to study the heavy-quark dynamics and find new physics(NP) beyond Standard Model(SM).
\end{abstract}


\maketitle

\section{Introduction}\label{sec1}

Since the pesudoscalar meson $B_c$ was firstly observed in 1998\cite{CDF:1998axz}, physicists have focused attention on it for its special properties. Because of the instability of top quark, $B_c$ meson is the only quarknoium with mixed heavy flavor quarks. Moreover, $B_c$ meson lies below the threshold of $B\bar D$, thus can only decay through weak interaction. Correspondingly, it is relatively more stable than other heavy quarkonium states consisting of only one flavor quark such as $\bar{c}c$ and $\bar{b}b$. Naturally, $B_c$ meson is an excellent laboratory to study the heavy-quark dynamics and the weak decay process about heavy quark.

The decay process of $B_c$ to charmonium is especially interesting and valuable. At quark level, the decay process that $B_c$ decays to charmonium is accompanied by $b \to c$ transition, which is useful to extract the Cabibbo-Kobayashi-Maskawa (CKM) matrix element $V_{cb}$. Furthermore, the decay process $B_c \to J/\psi \mathcal{l} \nu$ is similar to $B \to D^{(*)}\mathcal{l} \nu$. Both of these processes are relevant to the transition of $b \to c \mathcal{l} \nu$. As for the decay ratio $R(D^{(*)})=\Gamma(\bar{B} \to D^{(*)} \tau^- \bar{\nu})/\Gamma(\bar{B} \to D^{(*)} \mu^- \bar{\nu})$, the experimental results and predictions of the Standard Model (SM) were not agree with each other for a long time. Though Belle and LHCb carried out repeated measurements and theorists also updated their results for several times basing on SM predictions, there is still $3\sigma \sim 4\sigma$ discrepancy between them. The decay process $B_c \to J/\psi \mathcal{l} \nu$ which is similar to that of $B \to D^{(*)}\mathcal{l} \nu$, is expected to find the new physics (NP) beyond the Standard Model (SM). Recently, the LHCb collaboration measured the decay ratio\cite{LHCb:2017vlu},

\begin{align}
	R(J/\psi ) = \frac{{\Gamma (B_c^ -  \to J/\psi {\tau ^ - }\bar \nu )}}{{\Gamma (B_c^ -  \to J/\psi {\mu ^ - }\bar \nu )}} = 0.71 \pm 0.17(\mathrm{stat}) \pm 0.18(\mathrm{syst})
\end{align}
which waits for further confirmation by theorist.

Both experimental and theoretical investigations about $B_c$ decaying to charmonium have made great progress. The $B_c$ meson was firstly found through the decay process $B_c \to J/\psi \mathcal{l} \nu$ in $p\bar p$ collisions at the Fermilab Tevatron by the CDF collaboration. Up to now, many decay modes have been observed in experiments, such as $B_c^+ \to J/\psi \pi^+$\cite{LHCb:2012ihf}, $B_c^+ \to J/\psi \pi^+ \pi^- \pi^+$\cite{LHCb:2012ag}, $B_c^+ \to J/\psi K^+$\cite{LHCb:2013hwj}, $B_c^+ \to J/\psi D^{(*)+}_s$\cite{LHCb:2013kwl}, $B_c^+ \to J/\psi K^+ K^- \pi^+$\cite{LHCb:2013rud}. Thanks to the development of accelerator and detector techniques, it is now easy to reconstruct the decay process of $B_c \to J/\psi$. Ref.\cite{Gao:2010zzc} estimated that the inclusive production cross section of $B_c$ meson (including excited states) at the LHC is at a level of $1ub$ for $\sqrt{14}$ TeV. This means that $O(10^9)$ $B_c^+$ meson can be anticipated with 1 fb$^{-1}$, so more precise experimental results and more decay channels of $B_c$ are expected. As for the decay process $B_c \to \eta_c$, it has not been observed in experiments. In theory, the processes that $B_c$ decays to charmnoium have been studied by many theoretical methods, such as the lattice QCD\cite{Harrison:2020gvo, Colquhoun:2016osw}, the relativistic quark model (RQM)\cite{Nobes:2000pm,Faustov:2022ybm}, the covariant confined quarkmodel (CCQM)\cite{Tran:2018kuv},the non-relativistic QCD (NRQCD)\cite{Qiao:2012hp, Tang:2022nqm}, the light-front quark model (LFQM)\cite{Zhang:2023ypl,Shi:2016gqt}, the QCD sum rules (QCDSR)\cite{Kiselev:2000pp, Azizi:2013zta,Azizi:2019aaf,Colangelo:1992cx}, the light-cone QCD sum rules (LCSR)\cite{Wang:2007fs, Huang:2007kb,Leljak:2019eyw,Bordone:2022drp} and the perturbative QCD (pQCD)\cite{Wang:2012lrc,Liu:2023kxr,Liu:2020upy,Liu:2018kuo}. Nevertheless, these results are not consistent well with the experimental data or with each other, thus it is necessary for us to perform further confirmations about this weak transition by different theoretical methods.

The decay process of $B_c$ to charmonium is realized according to the spectator mechanism that $b$ quark decays and $c$ quark plays as a spectator. This process is related to the weak interaction and its effective Hamiltonian at quark level has been well formulated. The last step in calculating the decay width of $B_c$ decaying to charmonium is properly evaluating the hadron transition matrix element, which is commonly factorized as form factors. Generally, the form factors are non-perturbative in QCD, which can be obtained by non-perturbative methods. The QCDSR is a powerful non-perturbative approach which was proposed in Refs. \cite{Shifman:1978bx, Shifman:1978by}. Besides of the evaluation about hadron mass spectra and decay constants\cite{Wei:2006wa, Wang:2012kw, Aliev:2012iv, Aliev:2012tt}, QCDSR has been also used to analyze the electroweak and electromagnetic form factor\cite{Belyaev:1993wp, Dai:1996xv, Yang:2005bv, Shi:2019hbf, Zhao:2020mod, Zhang:2023nxl, Lu:2024tgy}, coupling constant of the strong interaction\cite{Bracco:2011pg, Yu:2015xwa, Lu:2023gmd, Lu:2023pcg, Lu:2023lvu,Colangelo:2000dp}, etc. In this work, we will calculate the form factors of $B_c$ decaying to $\eta_c$, $J/\psi$ in the framework of QCDSR with unified input parameters. In addition, these form factors will be used to obtain the decay widths of different decay processes including nonleptonic decays of $B_c^- \to \eta_c \pi^-$, $\eta_c K^-$, $\eta_c \rho^-$, $\eta_c K^{*-}$, $B_c^- \to J/\psi \pi^-$, $J/\psi K^-$, $J/\psi \rho^-$, $J/\psi K^{*-}$, and semileptonic decays of $B_c^- \to \eta_c \mathcal{l} \bar{\nu}$, $B_c^- \to J/\psi \mathcal{l} \bar{\nu}$.

The article is organized as follows. After introduction in Sec. \ref{sec1}, the basic procedure of three-point QCDSR are discussed in detail in Sec. \ref{sec2}. The numerical results of form factors are organized in Sec. \ref{sec3}. The calculation and discussion about decay width are performed in Sec. \ref{sec4}. In the last section a brief conclusion is presented.

\section{The three-point QCDSR}\label{sec2}

In the framework of QCDSR, the analysis about form factors starts from the three-point correlation function,
\begin{align}\label{eq:1}
\notag
\Pi (p,p') &= {i^2}\int {{d^4}x{d^4}y{e^{ip' \cdot x}}{e^{i(p - p') \cdot y}}} \\
&\times\left\langle 0 \right|T\{ {J_{C}}(x)J(y)J_{{B_c}}^ + (0)\} \left| 0 \right\rangle
\end{align}
where $T$ is the time ordering operation, $p$ and $p'$ are the momentum of $B_c$ meson and $C$ meson with $C=\eta_c$, $J/\psi$. $J_{C}$ and $J_{B_c}$ are interpolating currents which have the same quantum numbers with corresponding mesons. In this work, the scalar, vector and tensor form factors of $B_c \to \eta_c$, and vector, axial vector and tensor form factors of $B_c \to J/\psi$ are evaluated. Different form factors are distinguished from each other by the current $J(y)$. These currents in Eq. (\ref{eq:1}) can be written as,
\begin{align}\label{eq:2}
	\notag
	&{J_{{B_c}}}(0) = \bar c(0)i{\gamma _5}b(0)\\
	\notag
	&J(y) = \bar c(y)\Gamma b(y)\\
	\notag
	&{J_{{\eta _c}}}(x) = \bar c(x)i{\gamma _5}c(x)\\
	&{J^{J/\psi }_\mu}(x) = \bar c(x){\gamma _\mu }c(x)
\end{align}
where $\Gamma$ is the Dirac matrix, $\Gamma  = I,{\gamma _\mu },{\gamma _\mu }{\gamma _5},{\sigma _{\mu \nu }(\gamma_5)}$ for scalar, vector, axial vector and tensor form factors, respectively.

The correlation function will be calculated in two sides which are named as phenomenological side and QCD side. The former is realized according to phenomenological method where the form factors are extracted, whereas the latter is obtained by using the operations product expansion (OPE) in deep Euclidean region. Combining these two sides by the quark-hadron duality, the sum rules of form factors will be derived.

\subsection{\label{sec:level2.1}The Phenomenological Side}

In phenomenological side, the correlation function is inserted with complete sets of hadronic states with the same quantum numbers as $\eta_c$, $J/\psi$ and $B_c$. Performing integrations over $x$ and $y$, and using the dispersion relation, we can represent the correlation function of Eq. (\ref{eq:1}) as fallows\cite{Colangelo:2000dp},

\begin{align}\label{CFhadron}
	\notag
	\Pi^{\mathrm {had}1} (p,p') &= \frac{{\left\langle 0 \right|{J_{{\eta _c}}}(0)\left| {{\eta _c}(p')} \right\rangle \left\langle {{B_c}(p)} \right|J_{{B_c}}^ + (0)\left| 0 \right\rangle }}{{(m_{{B_c}}^2 - {p^2})(m_{{\eta _c}}^2 - {{p'}^2})}}\\
	\notag
	&\quad \times \left\langle {{\eta _c}(p')} \right|\bar c(0)b(0)\left| {{B_c}(p)} \right\rangle  + h.r.\\
	\notag
	\Pi^{\mathrm {had}2}_\mu (p,p') &= \frac{{\left\langle 0 \right|{J_{{\eta _c}}}(0)\left| {{\eta _c}(p')} \right\rangle \left\langle {{B_c}(p)} \right|J_{{B_c}}^ + (0)\left| 0 \right\rangle }}{{(m_{{B_c}}^2 - {p^2})(m_{{\eta _c}}^2 - {{p'}^2})}}\\
	\notag
	&\quad \times \left\langle {{\eta _c}(p')} \right|\bar c(0){\gamma _\mu }\left| {{B_c}(p)} \right\rangle  + h.r.\\
	\notag
	\Pi^{\mathrm {had}3}_{\mu\nu} (p,p') &= \frac{{\left\langle 0 \right|{J_{{\eta _c}}}(0)\left| {{\eta _c}(p')} \right\rangle \left\langle {{B_c}(p)} \right|J_{{B_c}}^ + (0)\left| 0 \right\rangle }}{{(m_{{B_c}}^2 - {p^2})(m_{{\eta _c}}^2 - {{p'}^2})}}\\
	\notag
	&\quad \times \left\langle {{\eta _c}(p')} \right|\bar c(0){\sigma _{\mu \nu }}{\gamma _5}b(0)\left| {{B_c}(p)} \right\rangle  + h.r.\\
	\notag
	\Pi^{\mathrm {had}4}_{\lambda\mu} (p,p') &= \frac{{\left\langle 0 \right|{J^{{J/\psi }}_\lambda}(0)\left| {{J/\psi }(p')} \right\rangle \left\langle {{B_c}(p)} \right|J_{{B_c}}^ + (0)\left| 0 \right\rangle }}{{(m_{{B_c}}^2 - {p^2})(m_{{J/\psi }}^2 - {{p'}^2})}}\\
	\notag
	&\quad \times \left\langle {{J/\psi}(p')} \right|\bar c(0){\gamma _\mu }b(0)\left| {{B_c}(p)} \right\rangle  + h.r.\\
	\notag
	\Pi^{\mathrm {had}5}_{\lambda\mu} (p,p') &= \frac{{\left\langle 0 \right|{J^{{J/\psi }}_\lambda}(0)\left| {{J/\psi }(p')} \right\rangle \left\langle {{B_c}(p)} \right|J_{{B_c}}^ + (0)\left| 0 \right\rangle }}{{(m_{{B_c}}^2 - {p^2})(m_{{J/\psi }}^2 - {{p'}^2})}}\\
	\notag
	&\quad \times \left\langle {{J/\psi}(p')} \right|\bar c(0){\gamma _\mu }{\gamma _5}b(0)\left| {{B_c}(p)} \right\rangle  + h.r.	\\
	\notag
	\Pi^{\mathrm {had}6}_{\lambda\mu\nu} (p,p') &= \frac{{\left\langle 0 \right|{J^{{J/\psi }}_\lambda}(0)\left| {{J/\psi }(p')} \right\rangle \left\langle {{B_c}(p)} \right|J_{{B_c}}^ + (0)\left| 0 \right\rangle }}{{(m_{{B_c}}^2 - {p^2})(m_{{J/\psi }}^2 - {{p'}^2})}}\\
	&\quad \times \left\langle {{J/\psi}(p')} \right|\bar c(0){\sigma _{\mu \nu }}b(0)\left| {{B_c}(p)} \right\rangle  + h.r.	
\end{align}
where $h.r.$ denotes the contributions of excited and continuum states. $\Pi^{\mathrm {had}1}$, $\Pi^{\mathrm {had}2}_\mu$ and $\Pi^{\mathrm {had}3}_{\mu\nu}$ are used to analyze the scalar, vector, and tensor form factors of $B_c \to \eta_c$, and $\Pi^{\mathrm {had}4}_{\lambda\mu}$, $\Pi^{\mathrm {had}5}_{\lambda\mu}$ and $\Pi^{\mathrm {had}6}_{\lambda\mu\nu}$ are used for the analysis of vector, axial vector and tensor form factors of $B_c \to J/\psi$, respectively. The meson vacuum matrix elements in these above correlation functions can be factorized as,
\begin{align}\label{matrix}
	\notag
	&\left\langle {\rm{0}} \right|{J_{{\eta _c}}}(0)\left| {{\eta _c}(p')} \right\rangle  = \frac{{{f_{{\eta _c}}}m_{{\eta _c}}^2}}{{2{m_c}}}\\
	\notag
	&\left\langle {\rm{0}} \right|{J^{J/\psi }_\lambda}(0)\left| {J/\psi (p')} \right\rangle  = {{f_{J/\psi }}m_{J/\psi }\xi_\lambda}\\
	&\left\langle {{B_c}(p)} \right|J_{{B_c}}^ + (0)\left| 0 \right\rangle  = \frac{{{f_{{B_c}}}m_{{B_c}}^2}}{{{m_c} + {m_b}}}
\end{align}
where $f_{\eta_c}$, $f_{J/\psi}$ and $f_{B_c}$ are decay constants of corresponding mesons. $\xi_{\lambda}$ is the polarization vector of $J/\psi$ meson. The other transition matrix elements in Eq. (\ref{CFhadron}) can be expressed in terms of form factors\cite{Colangelo:2022lpy},

\begin{widetext}
	\begin{align}\label{matrix2}
		\notag
		\left\langle {{\eta _c}(p')} \right|\bar c(0)b(0)\left| {{B_c}(p)} \right\rangle  &= {f_S}({q^2})\\
		\notag
		\left\langle {{\eta _c}(p')} \right|\bar c(0){\gamma _\mu }b(0)\left| {{B_c}(p)} \right\rangle  &= {f_ + }({q^2})\left( {{p_\mu } + {{p'}_\mu } - \frac{{m_{{B_c}}^2 - m_{{\eta _c}}^2}}{{{q^2}}}{q_\mu }} \right) + {f_0}({q^2})\frac{{m_{{B_c}}^2 - m_{{\eta _c}}^2}}{{{q^2}}}{q_\mu }\\
		\notag
		\left\langle {{\eta _c}(p')} \right|\bar c(0){\sigma _{\mu \nu }}{\gamma _5}b(0)\left| {{B_c}(p)} \right\rangle  &=  - \frac{{2{f_T}({q^2})}}{{{m_{{B_c}}} + {m_{{\eta _c}}}}}{\varepsilon _{\mu \nu \alpha \beta }}{p^\alpha }{{p'}^\beta }\\
		\notag
		\left\langle {J/\psi (p',\xi)} \right|\bar c(0){\gamma _\mu }b(0)\left| {{B_c}(p)} \right\rangle  &=  \frac{{2V({q^2})}}{{{m_{{B_c}}} + {m_{J/\psi }}}}{\varepsilon _{\mu \nu \alpha \beta }}{\xi ^{ * \nu }}{p^\alpha }{{p'}^\beta }\\
		\notag
		\left\langle {J/\psi (p',\xi )} \right|\bar c(0){\gamma _\mu }{\gamma _5}b(0)\left| {{B_c}(p)} \right\rangle  &= i({m_{{B_c}}} + {m_{J/\psi }})\left( {\xi _\mu ^ *  - \frac{{({\xi ^ * } \cdot q)}}{{{q^2}}}{q_\mu }} \right){A_1}({q^2}) + i({\xi ^ * } \cdot q)\frac{{2{m_{J/\psi }}}}{{{q^2}}}{q_\mu }{A_0}({q^2})\\
		\notag
		&\quad - i\frac{{({\xi ^ * } \cdot q)}}{{{m_{{B_c}}} + {m_{J/\psi }}}}\left( {{{(p{\rm{ + }}p')}_\mu } - \frac{{m_{{B_c}}^2 - m_{J/\psi }^2}}{{{q^2}}}{q_\mu }} \right){A_2}({q^2}) \\
		\left\langle {J/\psi (p',\xi )} \right|\bar c(0){\sigma _{\mu \nu }}b(0)\left| {{B_c}(p)} \right\rangle  &= {T_0}({q^2})\frac{{{\xi ^ * } \cdot q}}{{{{({m_{{B_c}}} + {m_{J/\psi }})}^2}}}{\varepsilon _{\mu \nu \alpha \beta }}{p^\alpha }{{p'}^\beta } + {T_1}({q^2}){\varepsilon _{\mu \nu \alpha \beta }}{p^\alpha }{\xi ^{ * \beta }} + {T_2}({q^2}){\varepsilon _{\mu \nu \alpha \beta }}{{p'}^\alpha }{\xi ^{ * \beta }}
	\end{align}
\end{widetext}
where $f_S$, $f_{+}/f_{0}$ and $f_T$ represent the scalar, vector and tensor form factors of $B_c \to \eta_c$, and $V$, $A_{0}/A_{1}/A_{2}$ and $T_{0}/T_{1}/T_{2}$ denote the vector, axial vector and tensor form factors of $B_c \to J/\psi$, respectively. These form factors are the function of four-momentum squared of $q$ with $q=p-p'$. With Eqs. (\ref{matrix})$\sim$(\ref{matrix2}) and the following relation of polarization vector,
\begin{align}
	{\xi _\mu }\xi _\nu ^* =  - {g_{\mu \nu }} + \frac{{{{p'}_\mu }{{p'}_\nu }}}{{{{p'}^2}}}
\end{align}
we can obtain the expressions of the correlation functions at phenomenological side. Take the vector form factor of $B_c \to J/\psi$ as an example, it is written as,
\begin{align}\label{eq7}
	\notag
	{\Pi ^{\mathrm {had}4}_{\lambda\mu} }(p,p')=&2\frac{{{f_{{B_c}}}{f_{J/\psi }}m_{{B_c}}^2{m_{J/\psi }}}}{{({m_{{B_c}}} + {m_{J/\psi }})({m_c} + {m_b})}}V(q^{2})\\
	&\times \frac{{{\varepsilon _{\lambda \mu \alpha \beta }}{p^\alpha }{{p'}^\beta }}}{{(m_{J/\psi }^2 - {{p'}^2})(m_{{B_c}}^2 - {p^2})}}+h.r.
\end{align}
Finally, we take the change of variables $p^{2}\rightarrow -P^{2}$, $p^{\prime2}\rightarrow -P^{\prime2}$ and $q^{2}\rightarrow -Q^{2}$, and perform the double Borel transformations for $P^{2}$ and $P^{\prime 2}$. The correlation function is written as,
\begin{align}
	\notag
	{\mathscr B\mathscr B}[{\Pi ^{\mathrm {had4}}_{\lambda \mu} }(P,P')] = &2\frac{{{f_{{B_c}}}{f_{J/\psi }}m_{{B_c}}^2{m_{J/\psi }}{\varepsilon _{\lambda \mu \alpha \beta }}{p^\alpha }{{p'}^\beta }}}{{({m_{{B_c}}} + {m_{J/\psi }})({m_c} + {m_b})}}{V}(Q^{2})\\
	&\times \exp ( - \frac{{m_{{B_c}}^2}}{{{M^2}}} - \frac{{m_{J/\psi }^2}}{{k{M^2}}})+{\mathscr B\mathscr B}(h.r.)
\end{align}
where $M$ is the Borel parameter. Theoretically there are two parameters introduced by the double Borel transformations, we can relate one to the other with factor $k=m_{\eta_c}^2/m_{B_c}^2$\cite{Bracco:2011pg}.

\subsection{\label{sec:level2.2}The QCD Side}

The correlation functions which have been written as Eq. (\ref{CFhadron}) at phenomenological side, now can be represented as follows according to the Wick`s theorem at the QCD side,
\begin{align}\label{QCDC}
	\notag
	\Pi^{\mathrm{QCD}1} (p,p') &= \int {{d^4}x{d^4}y{e^{ip'x}}{e^{i(p - p')y}}} \\
	\notag
	&\quad \times Tr\{ {C}( - x){\gamma _5}{C}(x - y){B}(y){\gamma _5}\}
	\\
	\notag
	\Pi^{\mathrm{QCD}2}_{\nu} (p,p') &= \int {{d^4}x{d^4}y{e^{ip'x}}{e^{i(p - p')y}}} \\
	\notag
	&\quad \times Tr\{ {C}( - x){\gamma _5}{C}(x - y)\gamma_{\nu}{B}(y){\gamma _5}\}
	\\
	\notag
	\Pi^{\mathrm{QCD}3}_{\mu\nu} (p,p') &= \int {{d^4}x{d^4}y{e^{ip'x}}{e^{i(p - p')y}}} \\
	\notag
	&\quad \times Tr\{ {C}( - x){\gamma _5}{C}(x - y){\sigma_{\mu\nu}}{\gamma_{5}}{B}(y){\gamma _5}\}
	\\
	\notag
	\Pi^{\mathrm{QCD}4}_{\lambda\mu} (p,p') &= \int {{d^4}x{d^4}y{e^{ip'x}}{e^{i(p - p')y}}} \\
	\notag
	&\quad \times Tr\{ {C}( - x){\gamma _\lambda}{C}(x - y)\gamma_{\mu}(y){\gamma _5}\}
	\\
	\notag
	\Pi^{\mathrm{QCD}5}_{\lambda\mu} (p,p') &= \int {{d^4}x{d^4}y{e^{ip'x}}{e^{i(p - p')y}}} \\
	\notag
	&\quad \times Tr\{ {C}( - x){\gamma _\lambda}{C}(x - y)\gamma_{\mu}{\gamma_5}{B}(y){\gamma _5}\}
	\\
	\notag
	\Pi^{\mathrm{QCD}6}_{\lambda\mu\nu} (p,p') &= \int {{d^4}x{d^4}y{e^{ip'x}}{e^{i(p - p')y}}} \\
	&\quad \times Tr\{ {C}( - x){\gamma _\lambda}{C}(x - y){\sigma_{\mu\nu}}{B}(y){\gamma _5}\}
\end{align}
where $C(x)$ and $B(x)$ are the full propagator of $c$ and $b$ quarks, which have the following expression\cite{Reinders:1984sr},
\begin{widetext}
\begin{align}\label{propagator}
	\notag
	Q_{ij}(x) &= \frac{i}{{{{(2\pi )}^4}}}\int {{d^4}k{e^{ - ik \cdot x}}} \bigg\{ \frac{{{\delta _{ij}}}}{{\slashed k - {m_{Q}}}}- \frac{{{g_s}G_{\alpha \beta }^nt_{ij}^n}}{4}\frac{{{\sigma ^{\alpha \beta }}(\slashed k + {m_{Q}}) + (\slashed k + {m_{Q}}){\sigma ^{\alpha \beta }}}}{{{{({k^2} - m_{Q}^2)}^2}}} + \frac{{{g_s}{D_\alpha }G_{\beta \lambda }^nt_{ij}^n({f^{\lambda \beta a}} + {f^{\lambda \alpha \beta }})}}{{3{{({k^2} - m_{Q}^2)}^4}}}\\
	&\quad - \frac{{g_s^2{{({t^a}{t^b})}_{ij}}G_{\alpha \beta }^aG_{\mu \nu }^b({f^{\alpha \beta \mu \nu }} + {f^{\alpha \mu \beta \nu }} + {f^{\alpha \mu \nu \beta }})}}{{4{{({k^2} - m_{Q}^2)}^5}}} + \frac{{i{\delta _{ij}}\left\langle {f{G^3}} \right\rangle }}{{48}} \frac{{(\slashed k + {m_{Q}})[\slashed k({k^2} - 3m_{Q}^2) + 2{m_{Q}}(2{k^2} - m_{Q}^2)](\slashed k + {m_{Q}})}}{{{{({k^2} - m_{Q}^2)}^6}}} + ... \bigg\}  
\end{align}
\end{widetext}
Here the subscript $i$ and $j$ are color indices, and ${\sigma _{\alpha \beta }} = i[{\gamma _\alpha },{\gamma _\beta }]/2$. ${D_\alpha } = {\partial _\alpha } - i{g_s}G_\alpha ^n{t^n}$, $G_\alpha^n$ is the gluon field, ${t^n} = {\lambda ^n}/2$, and ${\lambda ^n}$ is the Gell-Mann matrix. ${f^{\lambda \beta \alpha }}$ and ${f^{\alpha \beta \mu \nu }}$ are defined as
\begin{align}\label{eq:11}
	\notag
	{f^{\lambda \alpha \beta }} = &(\slashed k + {m_{c(b)}}){\gamma ^\lambda }(\slashed k + {m_{c(b)}}){\gamma ^a}(\slashed k + {m_{c(b)}}) \\
	\notag
	&\times{\gamma ^\beta }(\slashed k + {m_{c(b)}})\\
	\notag
	{f^{\alpha \beta \mu \nu }} = &(\slashed k + {m_{c(b)}}){\gamma ^\alpha }(\slashed k + {m_{c(b)}}){\gamma ^\beta }(\slashed k + {m_{c(b)}}) \\
	&\times{\gamma ^\mu }(\slashed k + {m_{c(b)}}){\gamma ^\nu }(\slashed k + {m_{c(b)}})
\end{align}
Substituting with the full propagator in Eq. (\ref{QCDC}), the correlation functions can be decomposed into different tensor structures except for $\Pi^{\mathrm{QCD}1}$,
\begin{align}\label{eq:13}
	\notag
	&\Pi _\nu ^{\mathrm{QCD2}}(p,p') = {\Pi^{\mathrm{QCD2}}_1}(p,p'){p_\nu } + {\Pi ^{\mathrm{QCD2}}_2}(p,p'){{p'}_\nu }\\
	\notag
	&\Pi _{\mu \nu }^{\mathrm{QCD3}}(p,p') = {\Pi ^{\mathrm{QCD3}}_1}(p,p'){\varepsilon _{\mu \nu \alpha \beta }}{p^\alpha }{{p'}^\beta }\\
	\notag
	&\Pi _{\lambda \mu }^{\mathrm{QCD4}}(p,p') = {\Pi ^{\mathrm{QCD4}}_1}(p,p'){\varepsilon _{\lambda \mu \alpha \beta }}{p^\alpha }{{p'}^\beta }\\
	\notag
	&\Pi _{\lambda \mu }^{\mathrm{QCD5}}(p,p') = {\Pi ^{\mathrm{QCD5}}_1}(p,p'){g_{\lambda \mu }} \\
	\notag
	&\quad + {\Pi ^{\mathrm{QCD5}}_2}(p,p'){p_\lambda }{p_\mu } + {\Pi ^{\mathrm{QCD5}}_3}(p,p'){{p'}_\lambda }{{p'}_\mu }\\
	\notag
	&\quad + {\Pi ^{\mathrm{QCD5}}_4}(p,p'){p_\lambda }{{p'}_\mu } + {\Pi ^{\mathrm{QCD5}}_5}(p,p'){{p'}_\lambda }{p_\mu }\\
	\notag
	&\Pi _{\lambda \mu \nu }^{\mathrm{QCD6}}(p,p') \\
	\notag
	&\quad = {\Pi ^{\mathrm{QCD6}}_1}(p,p'){\varepsilon _{\lambda \mu \nu \alpha }}{p^\alpha } + {\Pi ^{\mathrm{QCD6}}_2}(p,p'){\varepsilon _{\lambda \mu \nu \alpha }}{{p'}^\alpha } \\
	\notag
	&\quad + {\Pi ^{\mathrm{QCD6}}_3}(p,p'){\varepsilon _{\mu \nu \alpha \beta }}{p^\alpha }{{p'}^\beta }{p_\lambda } + {\Pi ^{\mathrm{QCD6}}_4}(p,p'){\varepsilon _{\mu \nu \alpha \beta }}{p^\alpha }{{p'}^\beta }{{p'}_\lambda } \\
	\notag
	&\quad + {\Pi ^{\mathrm{QCD6}}_5}(p,p'){\varepsilon _{\lambda \nu \alpha \beta }}{p^\alpha }{{p'}^\beta }{{p}_\mu } + {\Pi ^{\mathrm{QCD6}}_6}(p,p'){\varepsilon _{\lambda \nu \alpha \beta }}{p^\alpha }{{p'}^\beta }{{p'}_\mu } \\
	&\quad + {\Pi ^{\mathrm{QCD6}}_7}(p,p'){\varepsilon _{\lambda \mu \alpha \beta }}{p^\alpha }{{p'}^\beta }{p_\nu } + {\Pi ^{\mathrm{QCD6}}_8}(p,p'){\varepsilon _{\lambda \mu \alpha \beta }}{p^\alpha }{{p'}^\beta }{{p'}_\nu }
\end{align}
In these above equations, the expression $\Pi^{\mathrm{QCD}j}_i$ in the right side of the equations without lorentz index is called the invariant amplitude. Theoretically, each structure can be used to analyze the form factors. No matter which structure is used, it is feasible as long as two conditions are satisfied for the QCDSR. These two conditions will be discussed in detail in Sec. \ref{sec3}. In Eq. (\ref{eq:13}), each invariant amplitude can be divided into two parts,
\begin{align}\label{Eq17}
	\Pi_i (p,p') = \Pi ^{\mathrm{pert}} (p,p')+\Pi ^\mathrm{{non-pert}} (p,p')
\end{align}
where the non-perturbative term includes the two gluons condensate $\left\langle {g_s^2{G^2}} \right\rangle$ and three gluons condensate $\left\langle {f{G^3}} \right\rangle$. The contributions of higher dimensions than $\left\langle {f{G^3}} \right\rangle$ are neglected because these contributions are small and further suppressed by $O(\alpha_s^2)$.

\begin{figure}
	\centering
	\includegraphics[width=6cm, trim=0 0 0 0, clip]{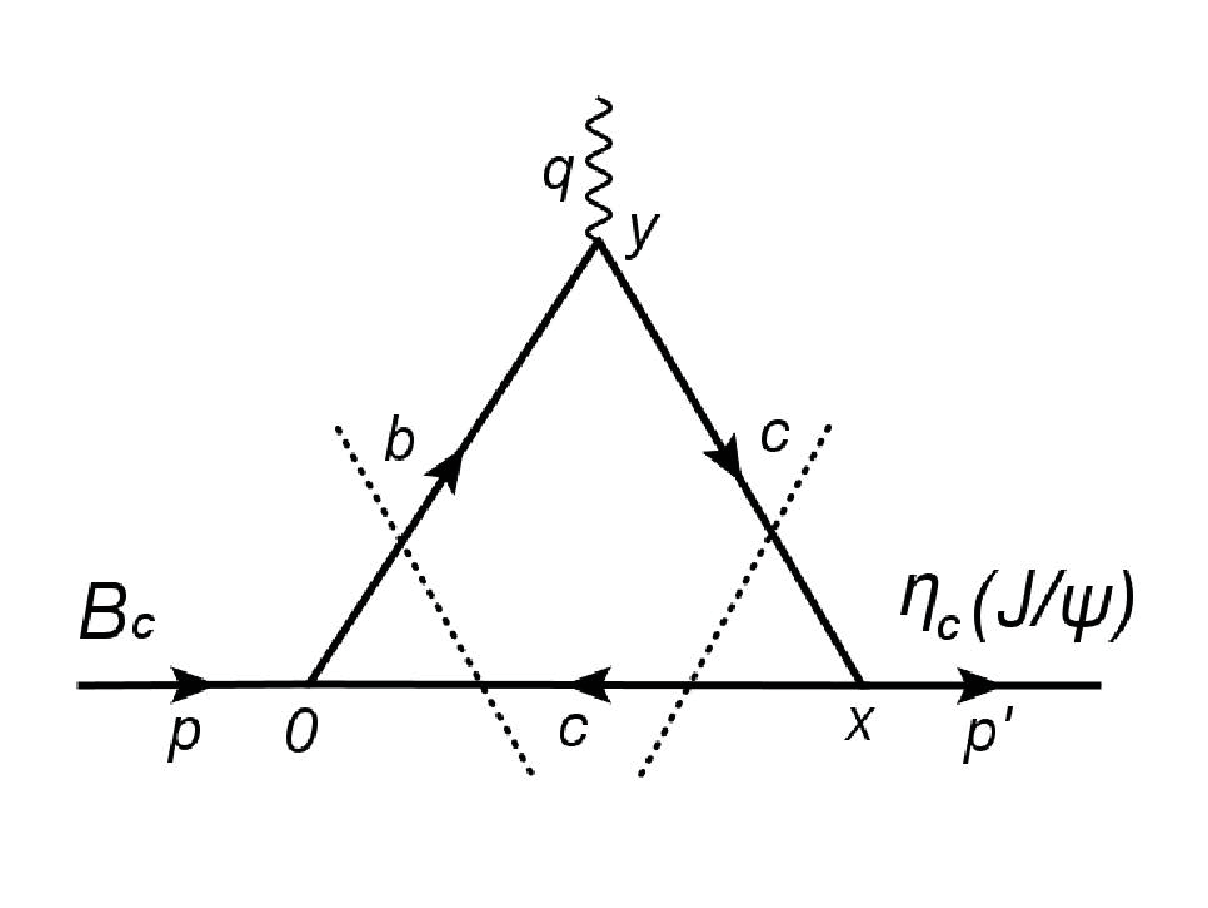}
	\caption{The feynman diagram for the perturbative part. The dashed lines denote the Cutkosky cuts.}
	\label{cutting rule}
\end{figure}

%
The next procedure is to obtain the expressions of spectral densities. Still taking the vector form factor $B_c \to J/\psi$ as an example, the perturbative part of the correlation function $\Pi _{\lambda \mu }^{\mathrm{QCD4}}$ in Eq. (\ref{QCDC}) can be written as,

\begin{align}\label{C0}
	\notag
	&\Pi _{\lambda \mu }^{\mathrm{QCD4pert}}(p,p') \\
	\notag
	&= \frac{{3{i^3}}}{{{{(2\pi )}^4}}}\int {{d^4}k} \frac{1}{{[{{(k - p')}^2} - m_c^2]({k^2} - m_c^2)[{{(k + p - p')}^2} - m_b^2]}}\\
	&\times Tr\{ [(\slashed k - \slashed p') + {m_c}]{\gamma _\lambda }(\slashed k + {m_c}){\gamma _\mu }[(\slashed k + \slashed p - \slashed p') + {m_b}]{\gamma _5}{\rm{\} }}
\end{align}
The spectral density can be derived according to the Cutkoskys's rules\cite{Cutkosky:1960sp} which is illustrated in Fig. \ref{cutting rule}. In this process, each denominator of quark propagator is substituted by a $\delta$ function,

\begin{figure*}[htbp]
	\centering
	\includegraphics[width=16cm]{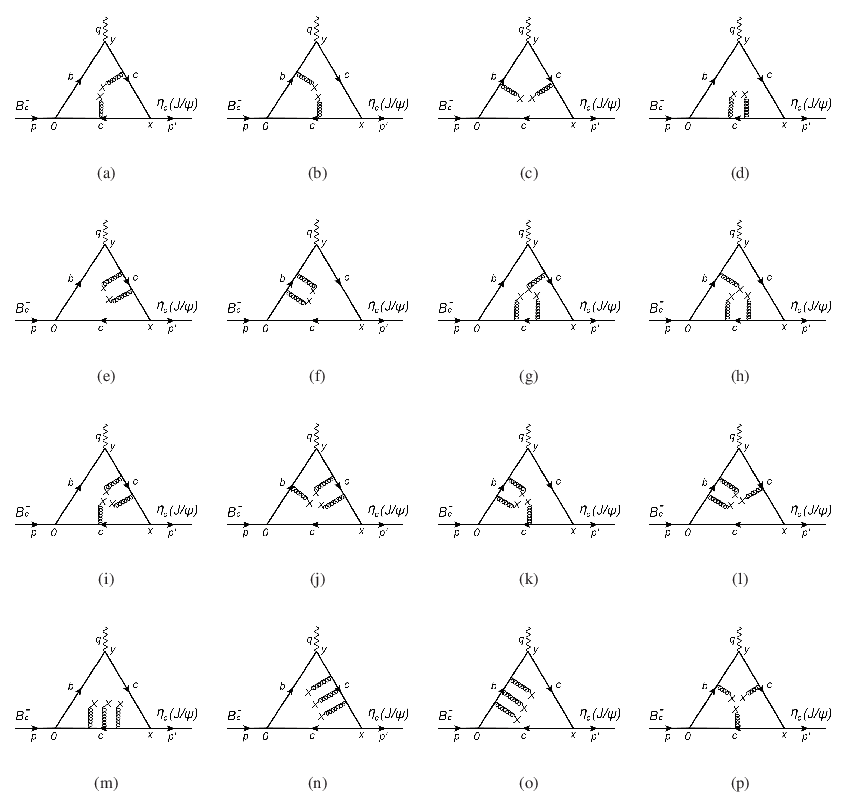}
	\caption{Feynman diagrams for the two-gluon (a-h) and three-gluon (i-p) condensate terms.}
	\label{feynman}
\end{figure*}
\begin{align}
	\frac{1}{{{k^2} - {m^2} + i\varepsilon }} \to  - 2\pi i\delta ({k^2} - {m^2})
\end{align}
Thus, there will be an integration for three delta functions, which can be derived as,
\begin{align}
	\notag
	&\int {{d^4}k\delta [{{(k - p')}^2} - m_c^2]\delta ({k^2} - m_b^2)\delta [{{(k + p - p')}^2} - m_c^2]} \\
	&= - \frac{\pi }{{2\sqrt {\lambda (s,u,{q^2})}}}
\end{align}
with
\begin{align}
\lambda (s,u,{q^2}) = {s^2} + {u^2} + {q^4} - 2(su + u{q^2} + s{q^2})
\end{align}
and $s=p^{2}$, $u=p^{\prime2}$. During this integral process, the following condition should be satisfied,
\begin{align}
	- 1 \le \frac{{2su - (m_b^2 - m_c^2 + u - {q^2})(s + u - {q^2})}}{{\sqrt {\left[ {{{(m_b^2 - m_c^2 + u - {q^2})}^2} - 4sm_c^2} \right]\lambda (s,u,{q^2})} }} \le 1
\end{align}
Finally, we obtain the spectral density of perturbative part,
\begin{align}
	\notag
	&\rho^\mathrm{{QCD4}}_{pert} (s,u,{q^2}) = \frac{3}{{4{\pi ^2}{{[{{(s + u-{q^2} )}^2} - 4su]}^{3/2}}}}\\
	\notag
	&\quad\times \{ {m_c}[ - 2m_b^2u + {q^4} - {q^2}(2s + u) + {s^2} - su]\\
	&\quad+ {m_b}u(2m_b^2 - {q^2} - s + u) - 2{m_b}m_c^2u + 2m_c^3u\}
\end{align}

As for the contributions of condensate terms, which are illustrated in Fig. \ref{feynman}, the calculating process is similar to that of the perturbative part. In this process, the following integration will be encountered,
\begin{align}
	{I_{abc}} = \int {{d^4}k\frac{1}{{{{[{{(k - p')}^2} - m_1^2]}^a}{{({k^2} - m_2^2)}^b}{{[{{(k + p - p')}^2} - m_3^2]}^c}}}}
\end{align}
By reducing the power of denominators, we can also perform the integration with the Cutkosky's rules,
\begin{align}
	\notag
	&{I_{abc}} = \frac{1}{{(a - 1)!(b - 1)!(c - 1)!}}\frac{{{\partial ^{a - 1}}}}{{\partial m_1^{a - 1}}}\frac{{{\partial ^{b - 1}}}}{{\partial m_2^{b - 1}}}\frac{{{\partial ^{c - 1}}}}{{\partial m_3^{c - 1}}}\\
	\notag
	&\times \int {{d^4}k\frac{1}{{[{{(k - p')}^2} - m_1^2]({k^2} - m_2^2)[{{(k + p - p')}^2} - m_3^2]}}} \\
	\notag
	&\to \frac{{{{( - 2\pi i)}^3}}}{{{{(2\pi i)}^2}}}\frac{1}{{(a - 1)!(b - 1)!(c - 1)!}}\frac{{{\partial ^{a - 1}}}}{{\partial m_1^{a - 1}}}\frac{{{\partial ^{b - 1}}}}{{\partial m_2^{b - 1}}}\frac{{{\partial ^{c - 1}}}}{{\partial m_3^{c - 1}}} \\
	&\times \frac{{ - \pi }}{{2\sqrt {\lambda (s,u,{q^2})} }}
\end{align}

After all of the spectral densities are obtained, the correlation function can be represented by spectral densities according to the dispersion relation,
\begin{align}
	\notag
	\Pi_{\lambda\mu}^{\mathrm{QCD4}} (p,p') = \int\limits_{s_{min}}^{\infty}  {\int\limits_{u_{min}}^{\infty}dsdu  {\frac{{{\rho }^{\mathrm{QCD4}}(s,u,{q^2})}}{{(s - {p^2})(u - p{'^2})}}{\varepsilon _{\lambda \mu \alpha \beta }}{p^\alpha }{{p'}^\beta }}}
\end{align}
where $s_{min} = (m_b + m_c)^2$ and $u_{min} = 4 m_c^2$. The Borel transformation can enhance the contribution of ground state and suppress those of excited and continuum states. After performing the double Borel transformations, we finally obtain the expression of correlation function at QCD side,
\begin{align}
	\notag
	&{\mathscr B\mathscr B}\Pi^{\mathrm{QCD4}}_{\lambda \mu} (M^{2},Q^{2}) = \int\limits_{{s_{min}}}^{s_{0}} {\int\limits_{{u_{min}}}^{u_{0}} {dsdu} {\rho}^{\mathrm{QCD4}} (s,u,{Q^2})}\\
	& \times \exp ( - \frac{s}{{{M^2}}} - \frac{u}{{k{M^2}}}){\varepsilon _{\lambda \mu \alpha \beta }}{p^\alpha }{{p'}^\beta }
\end{align}
where $s_0$ and $u_0$ are the threshold parameters. These two parameters are introduced to subtract the contributions of excited and continuum states.

After matching the phenomenological and QCD sides by using quark-hadron duality, the sum rule for the vector form factor of $B_c \to J/\psi$ is obtained,
\begin{align}
	\notag
	&{V}({Q^2}) = \int\limits_{{s_{min}}}^{{s_0}} {\int\limits_{{u_{min}}}^{{u_0}} {dsdu}{{{\rho }^{\mathrm{QCD4}}}(s,u,{Q^2})\exp ( - \frac{s}{{{M^2}}} - \frac{u}{{k{M^2}}})} }\\
	&\times \frac{{({m_{{B_c}}} + {m_{J/\psi }})({m_c} + {m_b})}}{{{\rm{2}}{f_{{B_c}}}m_{{B_c}}^2{f_{J/\psi }}{m_{J/\psi }}}}\exp (\frac{{m_{{B_c}}^2}}{{{M^2}}}{\rm{ + }}\frac{{m_{J/\psi }^2}}{{k{M^2}}})
\end{align}
As for the other form factors, they can also be obtained by similar process as that of ${V}({Q^2})$. All of the sum rules for other form factors are explicitly shown in Appendix \ref{A1}.

\begin{figure}[!]
	\centering
	\includegraphics[width=6cm, trim=0 0 0 0, clip]{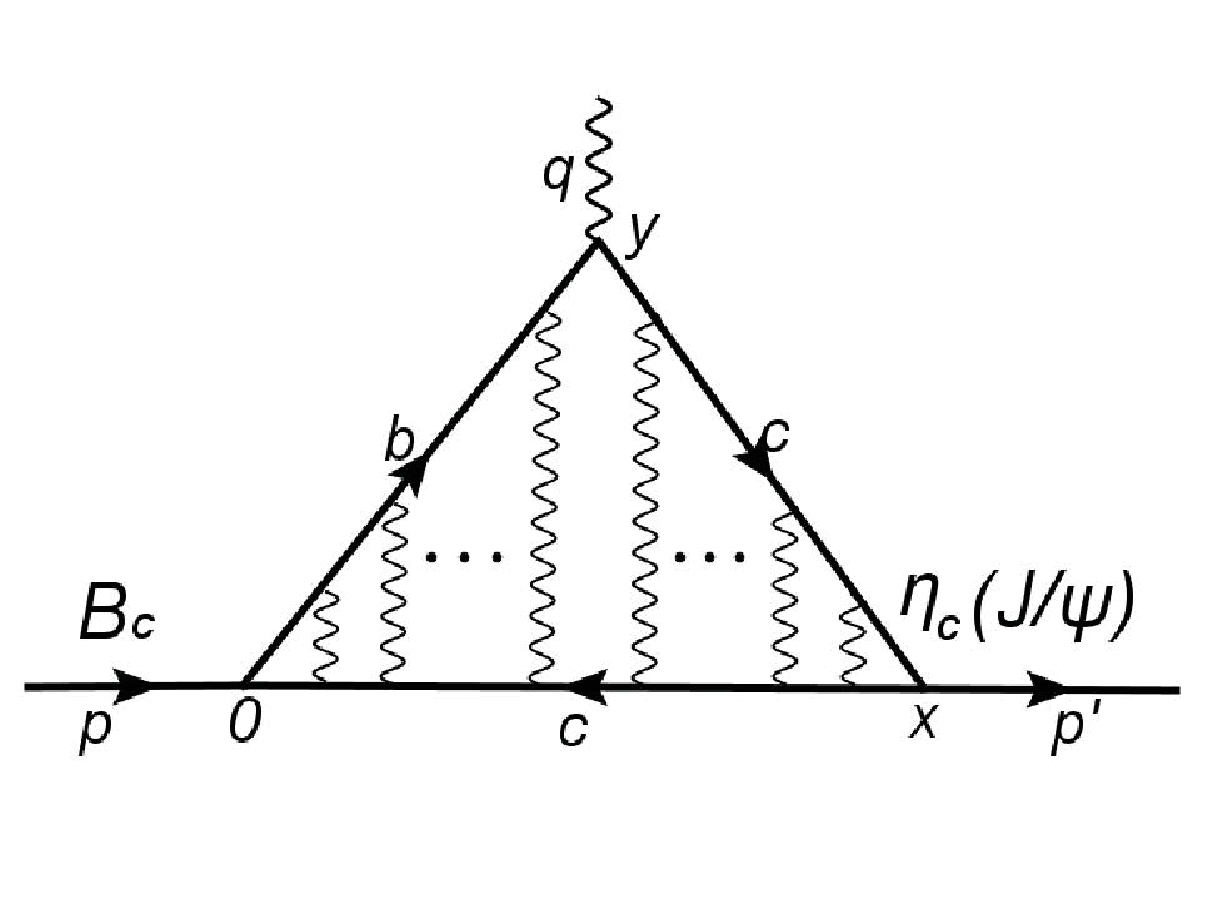}
	\caption{The ladder Feynman diagram for the Coulomb-like interaction.}
	\label{Coulomb-like}
\end{figure}

The Coulomb-like $\alpha_s/v$ correction is essential for heavy quarkonium $B_c$ meson and charmonium, where the relative velocity of quark movement is small. In this physical sketch,  the  Coulomb-like correction can be illustrated by the ladder diagram which is illustrated in Fig. \ref{Coulomb-like}. In the nonrelativistic approximation, this correction will lead to a renormalization coefficient for the spectral density of perturbative part\cite{Kiselev:1993ea, Kiselev:1999sc},
\begin{align}
	\rho^{pert}_c = \bm{C} \rho^{pert}
\end{align}
$\bm{C}$ is the correct coefficient with assumption $\alpha_s^C=\alpha_s(\mu)$,
\begin{align}
	\notag
	\bm{C} = &\sqrt {\frac{{4\pi \alpha _s^C}}{{3{v_1}}}{{\left[ {1 - \exp \left( { - \frac{{4\pi \alpha _s^C}}{{3{v_1}}}} \right)} \right]}^{ - 1}}}\\
	\times &\sqrt {\frac{{4\pi \alpha _s^C}}{{3{v_2}}}{{\left[ {1 - \exp \left( { - \frac{{4\pi \alpha _s^C}}{{3{v_2}}}} \right)} \right]}^{ - 1}}}
\end{align}
where $v_1$ and $v_2$ are the relative velocities of quarks in the $B_c$ meson and charmonium with
\begin{align}
	\notag
		{v_1} &= \sqrt {1 - \frac{{4{m_b}{m_c}}}{{{s_0} - {{({m_b} - {m_c})}^2}}}} \\
		{v_2} &= \sqrt {1 - \frac{{4m_c^2}}{{{u_0}}}}
\end{align}

\section{Numerical Results of form factors}\label{sec3}

The masses of quarks are energy-scale dependent, which can be expressed as the renormalization equation,
\begin{align}
	{m_{Q}}(\mu ) &= {m_{Q}}({m_{Q}}){\Big[\frac{{{\alpha _s}(\mu )}}{{{\alpha _s}({m_{Q}})}}\Big]^{\frac{{12}}{{33 - 2{n_f}}}}}\\
	{\alpha _s}(\mu ) &= \frac{1}{{{b_0}t}}\Big[1 - \frac{{{b_1}}}{{b_0^2}}\frac{{\log t}}{t}
	 + \frac{{b_1^2({{\log }^2}t - \log t - 1) + {b_0}{b_2}}}{{b_0^4{t^2}}}\Big]
\end{align}
where $t = \log \frac{{{\mu ^2}}}{{\Lambda _{QCD}^2}}$, ${b_0} = \frac{{33 - 2{n_f}}}{{12\pi }}$, ${b_1} = \frac{{153 - 19{n_f}}}{{24{\pi ^2}}}$, ${b_2} = \frac{{2857 - \frac{{5033}}{9}{n_f} + \frac{{325}}{{27}}n_f^2}}{{128{\pi ^3}}}$. $\Lambda _{QCD} = 213$ MeV for the flavors $n_f = 5$ in this work. The $\overline{\mathrm{MS}}$ masses are taken from the Particle Date Group (PDG)\cite{ParticleDataGroup:2022pth}, where $m_c(m_c)=1.275\pm0.025$ GeV and $m_b(m_b)=4.18\pm0.03$ GeV. In our previous studies\cite{Wang:2013iia,Wang:2013cha}, energy scale $\mu=2$ GeV worked well in analyzing the properties of $B_c$ meson, so this value is still employed in the present work. The values of gluons condensate are taken as the standard values from Refs. \cite{Narison:2010cg, Narison:2011xe, Narison:2011rn}. Threshold parameters $s_0$ and $u_0$ are used to eliminate the contributions of excited
and continuum states, which are commonly expressed as,
\begin{align}
	\notag
	{s_0} &= {({m_{{B_c}}} + \Delta )^2}\nonumber\\
	{u_0} &= {({m_C} + \Delta )^2}
\end{align}
Here the values of $s_0$ and $u_0$ should be larger than the ground state and lower than the first excited state. Commonly, the value of $\Delta$ with $0.4\sim0.6$ GeV can satisfy this condition. In present work, $\Delta=$0.5 GeV is used to obtain the central values of the final results, and 0.4, 0.6 GeV is for the lower and upper bounds, respectively\cite{Bracco:2011pg}. All the values of parameters used in this work are listed in Tab. \ref{parameters}.

\begin{table}
	\begin{ruledtabular}
		\caption{Input parameters used in calculation of form factors, the values with no reference have been mentioned in text. }
		\begin{tabular}{c c c c}
			Parameters & Values & Parameters & Values \\
			\hline
			$m_{B_c}$ & 6.27 GeV\cite{ParticleDataGroup:2022pth} & $f_{B_c}$ & 0.371 GeV\cite{ZGW:2024Bc,Narison:2020guz} \\
			$m_{\eta_c}$ & 2.98 GeV\cite{ParticleDataGroup:2022pth} & $f_{\eta_c}$ & 0.387 GeV\cite{Becirevic:2013bsa} \\
			$m_{J/\psi}$ & 3.10 GeV\cite{ParticleDataGroup:2022pth} & $f_{J/\psi}$ & 0.418 GeV\cite{Becirevic:2013bsa} \\
			$m_c(2 \mathrm{GeV})$ & 1.16 GeV & $\left\langle {g_s^2{G^2}} \right\rangle$ & $(0.88 \pm 0.15)$ GeV$^4$ \\
			$m_b(2 \mathrm{GeV})$ & 4.76 GeV & $\left\langle {g^{3}_{s}{GGG}f} \right\rangle $ & $(8.8 \pm 5.5)$ GeV$^2 \left\langle {g_s^2{GG}} \right\rangle$
			\label{parameters}
		\end{tabular}
	\end{ruledtabular}
\end{table}
To obtain stable and reliable results, two conditions should be satisfied for the QCDSR, which are the pole dominance and convergence of OPE.
The pole contribution is defined as,
\begin{align}
	\mathrm{Pole} = \frac{{\Pi _\mathrm{pole}({M^2})}}{{\Pi _\mathrm{total}({M^2})}}
\end{align}
with
\begin{align}
	\notag
	{\Pi _\mathrm{pole}}({M^2}) &= \int\limits_{{s_{min}}}^{{s_0}} {\int\limits_{{u_{min}}}^{{u_0}} {{\rho ^{QCD}}(s,u,{Q^2})\exp ( - \frac{s}{{{M^2}}} - \frac{u}{{k{M^2}}})dsdu} } \\
	{\Pi _\mathrm{total}}({M^2}) &= \int\limits_{{s_{min}}}^\infty  {\int\limits_{{u_{min}}}^\infty  {{\rho ^{QCD}}(s,u,{Q^2})\exp ( - \frac{s}{{{M^2}}} - \frac{u}{{k{M^2}}})dsdu} }
\end{align}
That is to say, we should find an appropriate region of the Borel parameters $M^{2}$, where the results have good stability and at the same time the condition of pole dominance($\geq40\%$) is also satisfied. This region is called the 'Borel platform'. Fixed $Q^2=1$ GeV$^2$, the Borel parameters are determined and Borel platform is chosen according to repeated trial and error. The variations of pole contribution, perturbative part and condensate terms with the Borel parameters are explicitly shown in Figs. \ref{PC} and \ref{BW} in Appendix \ref{A2}. From these figures, we can see that the condition of pole dominance is satisfied. At the same time, it is obvious that the contribution of perturbative term is dominant while those of gluon condensate terms are very small, which indicates the convergence of OPE. The Borel platform and the pole contribution are listed in Tab. \ref{numerical result of form factors}.
\begin{table}[htbp]
	\begin{ruledtabular}
		\caption{The Borel platform and pole contribution for different form factors}
		\begin{tabular}{c >{\centering}c c c }
			Modes & Form factors & Borel platforms & Pole contributions($\%$) \\
			\hline
			\multirow{4}{*}{$B_c \to \eta_c$}
			& $f_s$ & 14 $\sim$ 18 & 41.65\\
			& $f_+$ & 18 $\sim$ 22 & 40.91\\
			& $f_0$ & 18 $\sim$ 22 & 41.22\\
			& $f_T$ & 19 $\sim$ 23 & 40.24\\
			\hline
			\multirow{7}{*}{$B_c \to J/\psi$}
			& $V$ & 20 $\sim$ 24 & 40.14\\
			& $A_0$ & 16 $\sim$ 20 & 41.79\\
			& $A_1$ & 15 $\sim$ 19 & 41.89\\
			& $A_2$ & 13 $\sim$ 17 & 41.26\\
			& $T_0$ & 23 $\sim$ 27 & 40.98\\
			& $T_1$ & 19 $\sim$ 23 & 40.21\\
			& $T_2$ & 12 $\sim$ 16 & 40.74
			\label{numerical result of form factors}
		\end{tabular}
	\end{ruledtabular}
\end{table}
\begin{figure}[hbp]
	\centering
	\includegraphics[width=9cm]{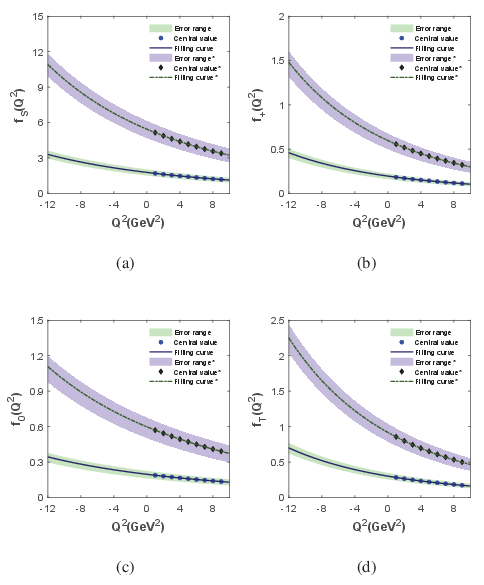}
	\caption{Fitting results of form factors for $B_c \to \eta_c$. The legends with diamond denote form factors with Coulomb-like correction.}
	\label{form factor etac}
\end{figure}
\begin{figure}[htbp]
	\centering
	\includegraphics[width=9cm, trim=0 0 0 0, clip]{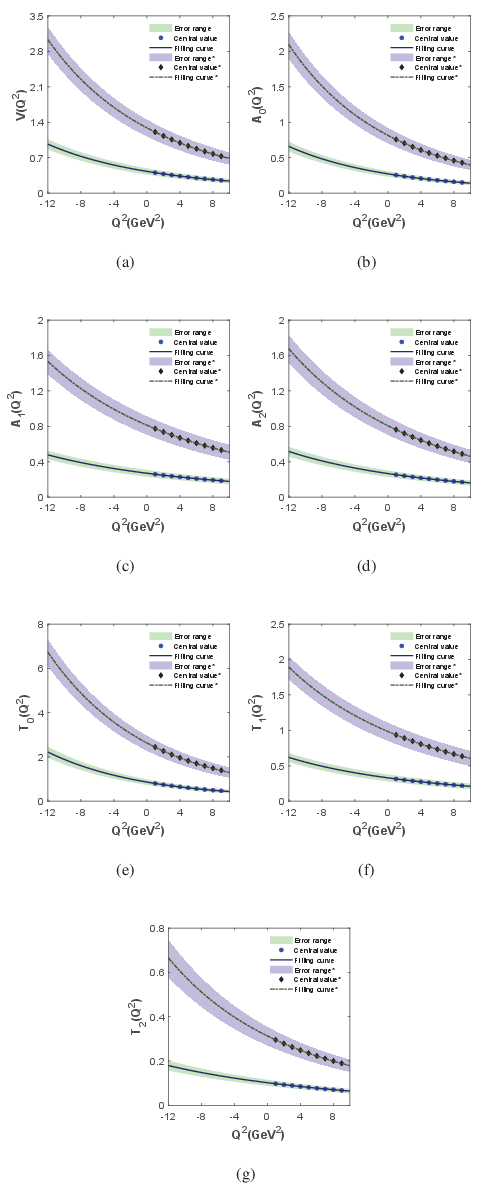}
	\caption{Fitting results of form factors for $B_c \to J/\psi$. The legends with diamond denote form factors with Coulomb-like correction.}
	\label{form factor Jpsi}
\end{figure}

By changing the value of $Q^{2}$, we obtain the form factors in space-like region ($1 \leq Q^2 \leq 9$), where the results are shown in Figs. \ref{form factor etac} and \ref{form factor Jpsi}. The values of form factors at $Q^2=0$ can be obtained by fitting the results in space-like region ($Q^{2}>0$) with appropriate analytical functions and then extrapolating these numerical results into the time-like region($Q^{2}<0$). The polar function, exponential function and the $z$-series parameterization approach are commonly used to fit the form factor\cite{Wang:2007ys, Wang:2008xt, Wang:2012vna,Biswas:2023bqz}. The $z$-series parameterization approach, which satisfies the unitarity, analyticity and perturbative QCD scaling, is adopted in this work. There are different forms of this approach used in different studies\cite{Boyd:1994tt, Bourrely:2008za, Wang:2015vgv}. Among them the method proposed in Ref. \cite{Bourrely:2008za} considered the asymptotic behaviors of the vector form factors near threshold. In this way form factors are expanded as,
\begin{align}\label{z}
	F({Q^2}) &= \frac{1}{{1 + {Q^2}/m_R^2}}\sum\limits_{k = 0}^{N - 1} {{b_k}[z{{({Q^2},{t_0})}^k} - {{( - 1)}^{k - N}}\frac{k}{N}z{{({Q^2},{t_0})}^N}]}
\end{align}
where $t_+ \equiv (m_{B_c}+m_C)^2$. $m_R$ denotes the masses of the low-lying $B_c$ resonances. In present work, $m_R$ is taken in accordance with Ref. \cite{Leljak:2019eyw}. $b_k$ are the real coefficients and $z(Q^2,t_0)$ is the function,
\begin{align}
	z({Q^2},{t_0}) &= \frac{{\sqrt {{t_ + } + {Q^2}}  - \sqrt {{t_ + } - {t_0}} }}{{\sqrt {{t_ + } + {Q^2}}  + \sqrt {{t_ + } - {t_0}} }}
\end{align}
which maps the complex $Q^2$-plane onto the unit disk $\lvert z(Q^2,t_0) \rvert \le 1$ in the $z$ complex plane. The free parameter $t_0$ corresponds to the value of $Q^2$ mapping onto the origin in the $z$-plane. Aimed to maximally reduce the interval of $z$ obtained after mapping of the semileptonic domain $Q^2 \in [-(m_{B_c}-m_C)^2,0]$, the auxiliary parameter $t_0$ are taken as\cite{Cui:2022zwm},
\begin{align}
	\notag
	{t_0} &= {t_ + } - \sqrt {{t_ + }({t_ + } - {t_ - })}\\
	t_- &= (m_{B_c}-m_C)^2
\end{align}
The $z$-series are truncated at $N=3$ in practical calculation. The vector and tensor form factors are expanded as Eq. (\ref{z}), while the scalar form factor can be well parameterized with a more simple function\cite{Bourrely:2008za, Cui:2022zwm},
\begin{align}
	f_S({Q^2}) = \frac{1}{{1 + {Q^2}/m_R^2}}\sum\limits_{k = 0}^{N - 1} {{b_k}[z{{({Q^2},{t_0})}^k}]}
\end{align}
\begin{table}[htbp]
\begin{ruledtabular}
	\caption{Fitting parameters of the $z$-series parameterized approach. $b_i^*$ denote the fitting parameters with Coulomb-like correction.}
	\begin{tabular}{ccccccccc}
		& Modes & Form factors & $b_0$ & $b_1$ & $b_2$ & $b_0^*$ & $b_1^*$ & $b_2^*$\\
		\hline
		& \multirow{4}{*}{$B_c \to \eta_c$} & $f_s$ & 2.1 & -17 & 37 & 6.5 & -67 & 22 \\
		&& $f_+$ & 0.25 & -3.3 & 15 & 0.78 & -11 & 55 \\
		&& $f_0$ & 0.22 & -1.5 & 1.4 & 0.69 & -5.6 & 7.7 \\
		&& $f_T$ & 0.38 & -4.9& 22 & 1.2 & -17 & 81 \\
		\hline
		& \multirow{7}{*}{$B_c \to J/\psi$} & $V$ & 0.52 & -6.6 & 30 & 1.6 & -22 & 100 \\
		&& $A_0$ & 0.34 & -4.9 & 24 & 1.1 & -17 & 90 \\
		&& $A_1$ & 0.30 & -2.2 & 3.8 & 0.94 & -8.5 & 22 \\
		&& $A_2$ & 0.31 & -3.0 & 7.9 & 0.97 & -11 & 40 \\
		&& $T_0$ & 1.1 & -20 & 140& 3.5 & -61 & 380 \\
		&& $T_1$ & 0.37 & -2.8 & 5.0 & 1.1 & -9.3 & 13 \\
		&& $T_2$ & 0.12 & -0.83 & -0.95 & 0.38 & -4.7 & 21
		\label{parameter}
	\end{tabular}
\end{ruledtabular}
\end{table}
\begin{table*}[htbp]
	\centering
	\renewcommand\arraystretch{1.5}
	\begin{ruledtabular}
		\caption{Form factors at $Q^2=0$, 'This work$^*$' denotes the results with Coulomb-like correction.}
		\label{compare}
		\begin{tabular}{c c c c c c c c c c}
			Modes & Form factors & This work & This work$^*$ & QCDSR\cite{Colangelo:1992cx} & Lattice QCD\cite{Harrison:2020gvo} & LCSR\cite{Leljak:2019eyw} & LFQM\cite{Zhang:2023ypl} & pQCD\cite{Wang:2012lrc} & RQM\cite{Nobes:2000pm}\\
			\hline
			\multirow{4}{*}{$B_c \to \eta_c$} & $f_S$ & $1.78^{+0.25}_{-0.26}$ & $5.44^{+0.65}_{-0.70}$ & - && - & - & - & - \\
			& $f_+$ & $0.20^{+0.03}_{-0.03}$ & $0.60^{+0.07}_{-0.09}$ & $0.20^{+0.02}_{-0.02}$ & - & $0.62^{+0.05}_{-0.05}$& 0.60 & $0.48^{+0.07}_{-0.07}$ & 0.54 \\
			& $f_0$ & $0.20^{+0.03}_{-0.03}$ & $0.60^{+0.07}_{-0.09}$ & - & - & - & 0.60 & $0.48^{+0.07}_{-0.07}$ & - \\
			& $f_T$ & $0.30^{+0.04}_{-0.05}$ & $0.92^{+0.11}_{-0.12}$ & - & - & $0.93^{+0.07}_{-0.07}$& - & - & - \\
			\hline
			\multirow{7}{*}{$B_c \to J/\psi$} & $V$ & $0.43^{+0.06}_{-0.06}$ & $1.29^{+0.15}_{-0.16}$ & $0.19^{+0.05}_{-0.05}$ & 0.725 & $0.73^{+0.06}_{-0.06}$ & 0.76 & $0.42^{+0.02}_{-0.02}$  & 0.73 \\
			& $A_0$ & $0.27^{+0.04}_{-0.04}$ & $0.82^{+0.1}_{-0.1}$ & - & 0.477 & $0.54^{+0.04}_{-0.04}$ & 0.55 & $0.59^{+0.03}_{-0.03}$ & 0.53 \\
			& $A_1$ & $0.27^{+0.04}_{-0.04}$& $0.81^{+0.09}_{-0.1}$ & $0.27^{+0.03}_{-0.03}$ & 0.457 & $0.55^{+0.04}_{-0.04}$ & 0.53 & $0.46^{+0.03}_{-0.03}$ & 0.52 \\
			& $A_2$ & $0.27^{+0.03}_{-0.04}$& $0.81^{+0.09}_{-0.1}$ & $0.28^{+0.09}_{-0.09}$ & 0.417 & $0.35^{+0.03}_{-0.03}$ & 0.49 & $0.64^{+0.03}_{-0.03}$ & 0.51\\
			& $T_0$ & $0.86^{+0.12}_{-0.12}$ & $2.63^{+0.3}_{-0.32}$ & - &-&-&-&-&-\\
			& $T_1$ & $0.33^{+0.05}_{-0.05}$ & $0.98^{+0.11}_{-0.12}$ & - &-&-&-&-&-\\
			& $T_2$ & $0.10^{+0.01}_{-0.01}$ & $0.31^{+0.04}_{-0.04}$ & - &-&-&-&-&-\\
		\end{tabular}
	\end{ruledtabular}
\end{table*}
All the fitting parameters $b_k$ are listed in Tab. \ref{parameter}. The fitting diagrams about the form factors of $B_c \to \eta_c$ and $B_c \to J/\psi$ are explicitly shown in Figs. \ref{form factor etac} and \ref{form factor Jpsi}. From these figures, we can see that the numerical results are all perfectly fitted by the analytical functions, which indicates the reliability of the final results. In addition, these figures also show that the values of form factors decrease evidently with increase of four-momentum squared $Q^{2}$. This $Q^{2}$ dependence is consistent well with the results of other collaborations\cite{Zhang:2023ypl,Colangelo:1992cx,Leljak:2019eyw,Harrison:2020gvo}. The physical transition processes of $B_{c}\rightarrow J/\psi$ and $B_{c}\rightarrow\eta_{c}$ are relevant to the time-like form factors which can be obtained by setting $Q^{2}\leq0$ in analytical functions. These time-like form factors will be used to analyze the nonleptonic and semileptonic decay processes of $B_{c}$ meson in Sec. \ref{sec4}. As a contrast, the values of the form factors at $Q^{2}=0$ are all displayed in Tab. \ref{compare}. It is indicated that the values of bare form factors calculated in this work are consistent well with the results of Ref.\cite{Colangelo:1992cx}, but some results are relatively small than those of other collaborations. After considering the Coulomb-like $\alpha_s/v$ correction, the numerical results are about three times of those without correction, which indicates the Coulomb-like correction has a significant impact on the results.

\section{\label{sec4}Weak decays of $B_c$ to charmonium}
\subsection{\label{sec4.1}Nonleptonic decay processes}

With the results of form factors obtained in the above section, the decay widths and branching ratios of several weak decay processes about $B_c \to \eta_{c}$ and $B_c \to J/\psi$ are calculated with the factorization approach\cite{Mohammadi:2014rpa, Mohammadi:2023itb}. The effective Hamiltonian of several nonleptonic decay processes of $B_c$ to charmonium are expressed as follows,
\begin{align}
	\notag
	&{H^{B_c^- \to C \pi(\rho)^-}_{eff}} = \frac{{{G_F}}}{{\sqrt 2 }}{V_{cb}}V_{ud}^*{a_1}\bar c{\gamma _\mu }(1 - {\gamma _5})b\bar d{\gamma ^\mu }(1 - {\gamma _5})u \\
	&{H^{B_c^- \to C K^{(*)-}}_{eff}} = \frac{{{G_F}}}{{\sqrt 2 }}{V_{cb}}V_{us}^*{a_1}\bar c{\gamma _\mu }(1 - {\gamma _5})b\bar u{\gamma ^\mu }(1 - {\gamma _5})s
\end{align}
where $V_{cb}$, $V_{ud}$ and $V_{us}$ are CKM matrix elements, $C=\eta_{c}$, $J/\psi$, and $a_1$ is the Wilson coefficients. All of the parameters used in this section are listed in Tab. \ref{parameters2}
\begin{table}
	\begin{ruledtabular}
		\caption{Input parameters used to analyze the decay processes of $B_c$.}
		\begin{tabular}{c c c c}
			Parameters & Values & Parameters & Values \\
			\hline
			$m_{\pi}$ & 0.14 GeV\cite{ParticleDataGroup:2022pth} & $f_{\pi}$ & 0.131 GeV\cite{ParticleDataGroup:2020ssz} \\
			$m_{K}$ & 0.494 GeV\cite{ParticleDataGroup:2022pth} & $f_{K}$ & 0.160 GeV\cite{ParticleDataGroup:2020ssz} \\
			$m_{\rho}$ & 0.775 GeV\cite{ParticleDataGroup:2022pth} & $f_{\rho}$ & 0.216 GeV\cite{ParticleDataGroup:2010dbb} \\
			$m_{K^*}$ & 0.892 GeV\cite{ParticleDataGroup:2022pth} & $f_{K^*}$ & 0.217 GeV\cite{ParticleDataGroup:2020ssz} \\
			$V_{cb}$ & 0.041\cite{ParticleDataGroup:2022pth} & $V_{ud}$ & 0.974\cite{ParticleDataGroup:2022pth} \\
			$V_{us}$ & 0.224\cite{ParticleDataGroup:2022pth} & $a_1$ & 1.07\cite{Buchalla:1995vs}
			\label{parameters2}
		\end{tabular}
	\end{ruledtabular}
\end{table}

The decay amplitude can be explicitly expressed as,
\begin{align}\label{decay amplitude}
	\Gamma  = \frac{1}{{2J + 1}}\sum\limits_J { \frac{1}{{2{m_{{B_c}}}}}\int {d\Phi (p \to p',q)} |T{|^2}}
\end{align}
where $J$ is the angle momentum of $B_c$ meson. $d\Phi (p \to p',q)$ is the two-body phase factor,
\begin{align}
	d\Phi (p \to q,p') = {(2\pi )^4}{\delta ^4}(p - q - p')\frac{{{d^3}\vec q}}{{{{(2\pi )}^3}2{q_0}}}\frac{{{d^3}\vec p'}}{{{{(2\pi )}^3}2{p'_0}}}
\end{align}
where $p$, $p'$ and $q$ are the four-momentum of $B_c$ meson, charmoniums($\eta_{c}$, $J/\psi$) and the other final state mesons($\pi^{-}$, $K^{-}$, $\rho$, $K^{*}$). After performing integration of the two-body phase factor, we can obtain the following expression,
\begin{align}
	\int {d\Phi (p \to p',q) = } \frac{1}{{8\pi m_{{B_c}}^2}}\sqrt {\lambda (m_{{B_c}}^2,m_C^2,{m^2})}
\end{align}
where $m_C$ and $m$ are the masses of charmonium $m_{\eta_{c}}$, $m_{J/\psi}$, and the other final state mesons $m_{\pi^-}$, $m_{K^{-}}$ $m_{\rho}$, $m_{K^{*}}$. $\lambda$ function can be expressed as,
\begin{align}
	\lambda (a,b,c) = {a^2} + {b^2} + {c^2} - 2(ab + bc + ac)
\end{align}

\begin{figure}[htbp]
	\centering
	\includegraphics[width=8cm]{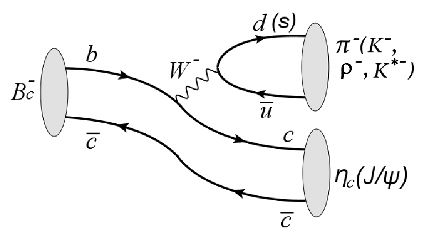}
	\caption{Feynman diagram for the decay processes $B_c^- \to \eta_{c} \pi^-, \eta_{c} K^-, \eta_{c} \rho^-, \eta_{c} K^{*-}$, and $B_c^- \to J/\psi \pi^-, J/\psi K^-, J/\psi \rho^-, J/\psi K^{*-}$}
	\label{factorization}
\end{figure}

In Eq. (\ref{decay amplitude}), $T$ is the transition matrix element. Under the factorization approach, this process can be illustrated as Fig. \ref{factorization}. Transition matrix element $T$ can be expressed as a product of two matrix elements,
\begin{align}
	\notag
	&\left\langle {{C}(p')\pi^- (q)} \right|{H_{eff}}\left| {{B_c^-}(p)} \right\rangle \\
	\notag
	&=\frac{{{G_F}}}{{\sqrt 2 }}{V_{cb}}V_{ud}^*{a_1}\left\langle {{C}(p')} \right|\bar c{\gamma _\mu }(1 - {\gamma _5})b\left| {{B^-_c}(p)} \right\rangle \\
	\notag
	&\quad \times \left\langle {\pi^- (q)} \right|\bar d{\gamma ^\mu }(1 - {\gamma _5})u\left| 0 \right\rangle
\end{align}
\begin{align}
	\notag
	&\left\langle {{C}(p')K^- (q)} \right|{H_{eff}}\left| {{B_c^-}(p)} \right\rangle \\
	\notag
	&=\frac{{{G_F}}}{{\sqrt 2 }}{V_{cb}}V_{us}^*{a_1}\left\langle {{C}(p')} \right|\bar c{\gamma _\mu }(1 - {\gamma _5})b\left| {{B^-_c}(p)} \right\rangle \\
	\notag
	&\quad \times \left\langle {K^- (q)} \right|\bar u{\gamma ^\mu }(1 - {\gamma _5})s\left| 0 \right\rangle
\end{align}
\begin{align}
	\notag
	&\left\langle {{C}(p')\rho^- (q)} \right|{H_{eff}}\left| {{B_c^-}(p)} \right\rangle \\
	\notag
	&=\frac{{{G_F}}}{{\sqrt 2 }}{V_{cb}}V_{ud}^*{a_1}\left\langle {{C}(p')} \right|\bar c{\gamma _\mu }(1 - {\gamma _5})b\left| {{B^-_c}(p)} \right\rangle \\
	\notag
	&\quad \times \left\langle {\rho^- (q)} \right|\bar c{\gamma ^\mu }(1 - {\gamma _5})s\left| 0 \right\rangle
\end{align}
\begin{align}\label{two body}
	\notag
	&\left\langle {{C}(p')K^{*-} (q)} \right|{H_{eff}}\left| {{B_c^-}(p)} \right\rangle \\
	\notag
	&=\frac{{{G_F}}}{{\sqrt 2 }}{V_{cb}}V_{us}^*{a_1}\left\langle {{C}(p')} \right|\bar c{\gamma _\mu }(1 - {\gamma _5})b\left| {{B^-_c}(p)} \right\rangle \\
	&\quad \times \left\langle {K^{*-} (q)} \right|\bar c{\gamma ^\mu }(1 - {\gamma _5})s\left| 0 \right\rangle
\end{align}
where $C=\eta_{c}$, $J/\psi$.
In these above equations (Eq. (\ref{two body})), each matrix element of $B_{c}\rightarrow \eta_{c}$ or $B_{c}\rightarrow J/\psi$ can be factorized as the form factor which has been listed in Eq. (\ref{matrix2}). The other meson vacuum matrix elements can be factorized as decay constant. As for pseudoscalar meson $\pi^{-}$ and $K^{-}$, and axisvector meson $\rho^{-}$ and $K^{*-}$, the decay constants are respectively defined as,
\begin{align}
	\notag
	\left\langle {P(q)} \right|\bar q{\gamma _\mu }(1 - {\gamma _5})q\left| 0 \right\rangle  &=  - i{f_P}{q_\mu }\\
	\left\langle {A(q)} \right|\bar q{\gamma _\mu }(1 - {\gamma _5})q\left| 0 \right\rangle  &=  - i{f_A}{m_A}{\xi_\mu }
\end{align}
where $P = \pi^{-}$, $K^{-}$, and $A= \rho^{-}$, $K^{*-}$. Substituting these matrix elements in Eq. (\ref{decay amplitude}), the expression of decay width can be derived. All the expressions of decay widths are explicitly shown in Appendix \ref{A3}. Finally, the decay widths and branching ratios of the nonleptonic decays are calculated and the corresponding numerical results are explicitly shown in Tab. \ref{BR}.

\begin{table*}[htbp]
	\renewcommand\arraystretch{1.5}
	\begin{ruledtabular}
		\caption{Decay widths (in $10^{-7}$eV) and Branching ratios (in \perthousand) of $B_c$ decaying to charmonium. Branching ratios are calculated at $\tau_{B_c}$ = 0.51 ps\cite{LHCb:2014ilr}. The superscript star denotes the results obtained by considering Coulomb-like correction.}
		\label{BR}
		\begin{tabular}{c >{\centering}p{7em}| c c | c c c c c c c}
			&\multirow{2}{*}{Decay channels} & \multicolumn{2}{c|}{Decay widths} & \multicolumn{7}{c}{Branching ratios}\\
			&&This work & This work$^*$ &This work & This work$^*$ & QCDSR\cite{Kiselev:2002vz} & LFQM\cite{Zhang:2023ypl} & pQCD\cite{Wang:2012lrc} & RQM\cite{Ebert:2003cn} & LCSR\cite{Leljak:2019eyw}\\
			\hline
			&$B_c^- \to \eta_c \pi^-$ & $1.8^{+0.6}_{-0.6}$ & $17^{+5}_{-5}$ & $0.15^{+0.05}_{-0.05}$ &  $1.4^{+0.4}_{-0.4}$ & 2 & $2.36 ^{+0.1}_{-0.1}$ & - & 0.85 & -\\
			&$B_c^- \to \eta_c K^-$ & $0.15^{+0.05}_{-0.05}$ & $1.4^{+0.4}_{-0.4}$ & $0.012^{+0.003}_{-0.003}$ &  $0.11^{+0.03}_{-0.03}$ & 0.13 & $0.19^{+0}_{-0.01}$ & - & 0.07 & -\\
			&$B_c^- \to \eta_c \rho^-$ & $4.9^{+1.5}_{-1.5}$ & $46^{+12}_{-13}$ & $0.41^{+0.12}_{-0.12}$  & $3.8^{+0.1}_{-0.1}$ & 4.2 & $3.8^{+1.0}_{-1.0}$ & - & 2.1 & -\\
			&$B_c^- \to \eta_c K^{*-}$ & $0.26^{+0.08}_{-0.08}$ & $2.5^{+0.6}_{-0.7}$ & $0.022^{+0.007}_{-0.007}$ &  $0.21^{+0.05}_{-0.06}$ & 0.20 & $0.34^{+0.02}_{-0.02}$ & - & 0.11 & -\\
			&$B_c^- \to \eta_c e \bar{\nu}_e$ & $11^{+3}_{-3}$ & $105^{+23}_{-26}$ & $0.9^{+0.2}_{-0.3}$  & $8.7^{+1.9}_{-2.2}$ & 7.5 & - & 4.41 & 4.2 & 8.2\\
			&$B_c^- \to \eta_c \mu \bar{\nu}_{\mu}$ & $11^{+3}_{-3}$ & $105^{+23}_{-26}$  & $0.9^{+0.2}_{-0.3}$  & $8.7^{+1.9}_{-2.2}$ & - & - & 4.41 & - & 8.2\\
			&$B_c^- \to \eta_c \tau \bar{\nu}_{\tau}$ & $3.3^{+0.8}_{-0.9}$ & $33^{+7}_{-8}$ & $0.28^{+0.07}_{-0.07}$ & $2.8^{+0.5}_{-0.6}$ & 2.3 & - & 1.37 & - & 2.6\\
			&$B_c^- \to J/\psi \pi^-$ & $3.3^{+1}_{-0.9}$ & $30^{+8}_{-7}$ & $0.27^{+0.08}_{-0.07}$ & $2.5^{+0.6}_{-0.6}$ & 1.3 & $1.97^{+0.28}_{-0.29}$ & - & 0.61 & -\\
			&$B_c^- \to J/\psi K^-$ & $0.26^{+0.07}_{-0.07}$ & $2.3^{+0.6}_{-0.5}$ & $0.022^{+0.006}_{-0.006}$  & $0.19^{+0.05}_{-0.04}$ & 0.11 & $0.16^{+0.02}_{-0.02}$ & - & 0.05 & -\\
			&$B_c^- \to J/\psi \rho^-$ & $11^{+2.8}_{-3.0}$ & $96^{+22}_{-23}$ & $0.88^{+0.23}_{-0.25}$ & $8.0^{+1.9}_{-2.0}$ & 4.0 & $5.34^{+0.16}_{-0.16}$ & - & 1.6 & -\\
			&$B_c^- \to J/\psi K^{*-}$ & $0.60^{+0.16}_{-0.17}$ & $5.4^{+1.2}_{-1.3}$ & $0.049^{+0.013}_{-0.014}$  & $0.45^{+0.10}_{-0.11}$ & 0.22 & $0.31^{+0.14}_{-0.14}$ & - & 0.10 & -\\
			&$B_c^- \to J/\psi e \bar{\nu}_e$ & $61^{+15}_{-14}$ & $554^{+150}_{-91}$ & $5.1^{+1.2}_{-1.2}$  & $46^{+12}_{-08} $ & 19 & - & 10.03 & 12.3 & 22.4\\
			&$B_c^- \to J/\psi \mu \bar{\nu}_{\mu}$ & $61^{+15}_{-14}$ & $552^{+149}_{-90}$ & $5.1^{+1.2}_{-1.2}$  & $46^{+12}_{-8}$ & - & - & 10.03 & - & 22.4\\
			&$B_c^- \to J/\psi \tau \bar{\nu}_{\tau}$ & $15^{+3.5}_{-3.4}$ & $144^{+35}_{-23}$ &  $1.3^{+0.3}_{-0.3}$ & $12^{+3}_{-2}$ & 4.8 & - & 2.92 & - & 5.3\\	
		\end{tabular}
	\end{ruledtabular}
\end{table*}
\subsection{\label{sec4.2}Semileptonic decay processes}
\begin{figure}[htbp]
	\centering
	\includegraphics[width=8cm]{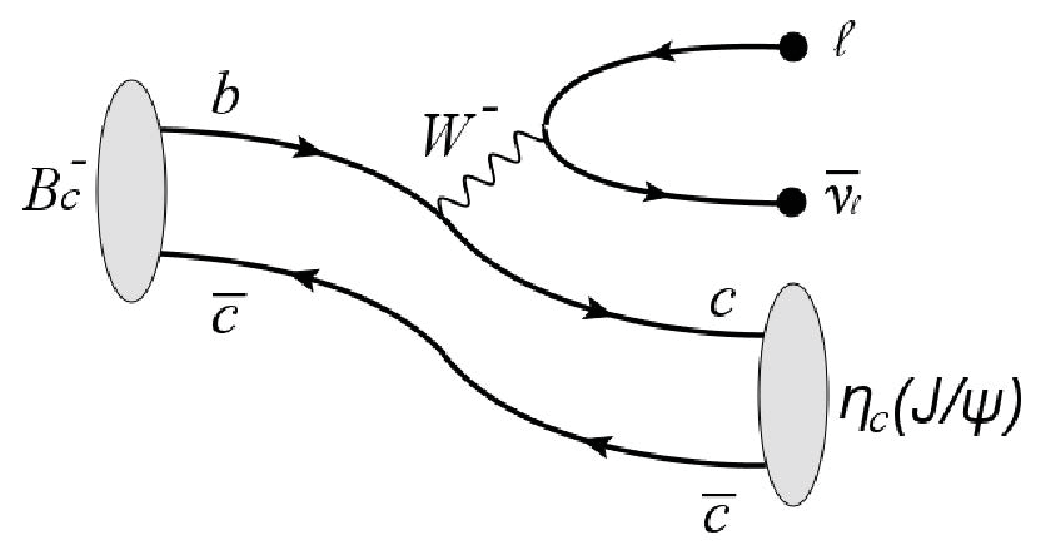}
	\caption{Feynman diagram for the decay processes $B_c^- \to J/\psi \mathcal{l} \bar{\nu}_{\mathcal{l}}$ and $B_c^- \to \eta_{c} \mathcal{l} \bar{\nu}_{\mathcal{l}}$.}
	\label{factorization2}
\end{figure}

\begin{figure}[htbp]
	\centering
	\includegraphics[width=8cm]{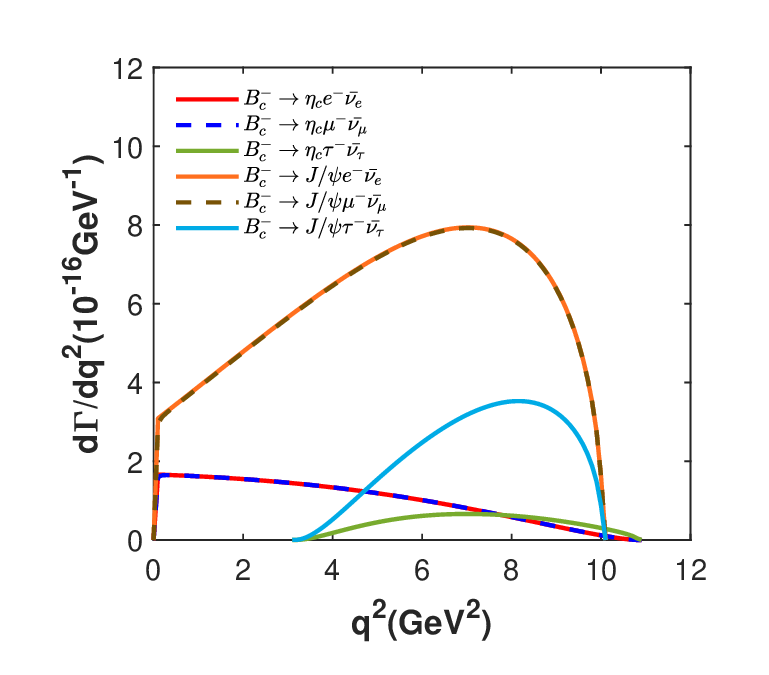}
	\caption{The differential decay widths $d\Gamma/dq^2$ with variations of $q^2$ for the processes $B_c^- \to C \mathcal{l} \bar{\nu}_{\mathcal{l}}$. These results are obtained with the bare form factors which are calculated without considering the Coulomb-like corrected.}
	\label{decay}
\end{figure}

\begin{figure}[htbp]
	\centering
	\includegraphics[width=8cm]{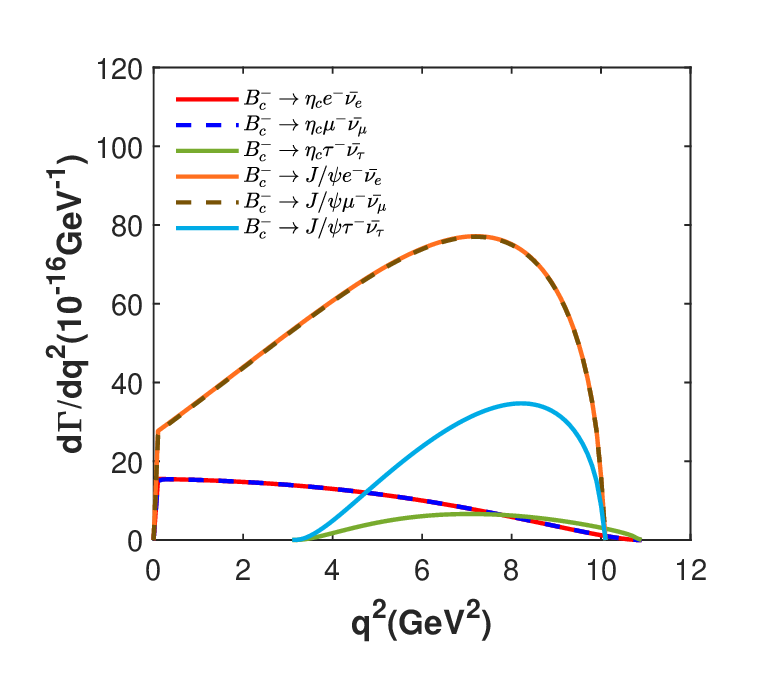}
	\caption{It is the same as Fig. 8, but these results are obtained with the form factors which are calculated by considering the Coulomb-like corrected.}
	\label{decayc}
\end{figure}

As for the semileptonic decay of $B_{c}$ meson, the effective Hamilton of the decay modes $B_c^- \to \eta_{c} \mathcal{l}^- \bar{\nu}_{\mathcal{l}}$ and $B_c^- \to J/\psi \mathcal{l}^- \bar{\nu}_{\mathcal{l}}$ can be expressed as,
\begin{align}
	H_{eff} = \frac{{{G_F}}}{{\sqrt 2 }}{V_{cb}}\bar c{\gamma _\mu }(1 - {\gamma _5})b{\bar \nu _l}{\gamma _\mu }(1 - {\gamma _5})l
\end{align}
The decay amplitude is the same as Eq. (\ref{decay amplitude}). The three-body phase factor can be decomposed as a product of double two-body phase factors,
\begin{align}
	\notag
	d\Phi(p \to p', l, \nu) &= {(2\pi )^4}{\delta ^4}(p - p' - l - \nu ) \\
	\notag
	&\quad \times \frac{{{d^3}\vec p'}}{{{{(2\pi )}^3}2{{p'}_0}}}\frac{{{d^3}\vec l}}{{{{(2\pi )}^3}2{l_0}}}\frac{{{d^3}\vec \nu}}{{{{(2\pi )}^3}2{\nu _0}}}\\
	&= \frac{{d{q^2}}}{{2\pi }}d\Phi (p \to p',q)d\Phi (q \to l,\nu )
\end{align}
Here $q$ is the four-momentum of $W^-$ boson, which is off shell.
This decay process can be illustrated as the feynman diagram in Fig.\ref {factorization2}, where the transition matrix element is decomposed into a product of $\left\langle {C} \right|\bar c{\gamma _\mu }(1 - {\gamma _5})b\left| {{B_c}} \right\rangle $ and $\left\langle {\ell \bar \nu } \right|{{\bar \nu }_l}{\gamma _\mu }(1 - {\gamma _5})l\left| 0 \right\rangle $ with $C=\eta_{c}$, $J/\psi$,
\begin{align}
	\notag
	&\left\langle {C(p'),{l }(l),\bar{\nu _l}(\nu)} \right|{H_{eff}}\left| {{B^-_c(p)}} \right\rangle \\
	\notag
	&= \frac{{{G_F}}}{{\sqrt 2 }}{V_{cb}}\left\langle {\ell \bar \nu }  \right|{{\bar \nu }_l}{\gamma _\mu }(1 - {\gamma _5})l\left|0 \right\rangle \\
	&\quad \times \left\langle {C(p')} \right|\bar c{\gamma _\mu }(1 - {\gamma _5})b\left| {{B^-_c(p)}} \right\rangle
\end{align}
It is similar as the two-body nonleptonic decay process, we can also derive the width of this semileptonic decay. The expressions of differential decay widths about these decay processes are listed in Appendix \ref{A4}. The differential decay widths with variations of $q^2$ are plotted in Figs. \ref{decay} and \ref{decayc}. The predictions for partial decay widths and branching ratios are also shown in Tab. \ref{BR}.

In most of the relevant references, only the branching ratios about these weak decay processes are shown explicitly. In this work, the differential decay widths, partial widths and the branching ratios are all listed explicitly. Comparing with other collaborations, the values of branching ratios without Coulomb-like $\alpha_s/v$ correction are lower than those of others. After considering the Coulomb-like $\alpha_s/v$ correction, the results are compatible with others, which indicates the importance of the correction. Using the values of form factors with or without Coulomb-like $\alpha_s/v$ correction, the ratio $\frac{{\Gamma (B_c^ -  \to J/\psi {\tau ^ - }\bar \nu_{\tau} )}}{{\Gamma (B_c^ -  \to J/\psi {\mu ^ - }\bar \nu_{\mu} )}}$ is the same with each other,
\begin{align}
	R(J/\psi ) = \frac{{\Gamma (B_c^ -  \to J/\psi {\tau ^ - }\bar \nu_{\tau} )}}{{\Gamma (B_c^ -  \to J/\psi {\mu ^ - }\bar \nu_{\mu} )}} = 0.25
\end{align}
This result is consistent well with that of other collaborations, where their results are in the range of $0.25\sim0.28$. However, the experimental data about this ratio is $0.71 \pm 0.17 \pm 0.18$, which is higher than the theoretical predictions. There are two possible interpretations about this deviation. First, the theoretical results have a certain model dependency. This deviation may be caused by the uncertainties of the experimental data and theoretical predictions. Besides, it is indicated by other researches that the similar problem also exist to the branching fraction $\frac{{\Gamma (B  \to D^{(*)} {\tau ^ - }\bar \nu_{\tau} )}}{{\Gamma (B  \to D^{(*)} {\mu ^ - }\bar \nu_{\mu} )}}$. Many theorists speculated that this kind of discrepancy may be the signs of new physics beyond the SM.

\begin{large}
\section{Conclusions}\label{sec5}
\end{large}

In this work, the form factors of $B_c \rightarrow\eta_c$ and $B_c \rightarrow J/\psi$ are calculated in the framework of three-point QCDSR. The Coulomb-like correction is considered, which indicates about triple the form factor at $Q^{2}=0$. All the form factors are obtained with the condition of pole dominance and convergence of OPE being well satisfied. We find that the values of form factors without Coulomb-like correction are in agreement well with the results of Ref.\cite{Colangelo:1992cx}, but lower than those of other theoretical methods. After considering the Coulomb-like correction, the results are compatible with those of other theoretical methods. Besides the form factors at $Q^{2}=0$, we also obtain their values of $Q^{2}>0$ by using the fitting function which is obtained with $z$-series parametrization approach. Using these values of form factors and basing on factorization approach, we also systematically analyze the decay widths and branching ratios of different decay channels. Our predictions are comparative with the results obtained by other theoretical approaches. However, the ratio $R(J/\psi)$ is somewhat lower than the experimental data, which indicates existence of the discrepancy between the prediction based on SM and experimental data. In summary, the results of form factors, decay widths and branching ratios obtained in this work provide useful information for further studying the heavy-quark dynamics and finding new physics beyond Standard Model.

\section*{Acknowledgements}

This project is supported by National Natural Science Foundation, Grant Number 12175068 and Natural Science Foundation of HeBei Province, Grant Number A2018502124.

\appendix

\begin{widetext}
	
\section{The expressions of form factors}\label{A1}
\begin{align}
	{f_S}({Q^2}) = \frac{{{m_b} + {m_c}}}{{{f_{{B_c}}}m_{{B_c}}^2}}\frac{{2{m_c}}}{{{f_{{\eta _c}}}m_{{\eta _c}}^2}}\int\limits_{{s_{\min }}}^{{s_0}} {\int\limits_{{u_{\min }}}^{{u_0}} {dsdu{\rho ^{\mathrm{QCD1}}}(s,u,{Q^2})} } \exp ( - \frac{s}{{{M^2}}} - \frac{u}{{k{M^2}}})\exp (\frac{{m_{{B_c}}^2}}{{{M^2}}} + \frac{{m_{{\eta _c}}^2}}{{k{M^2}}})
\end{align}
\begin{align}
	\notag
	\left( {\begin{array}{*{20}{c}}
			{{f_ + }({Q^2})}\\
			{{f_0}({Q^2})}
	\end{array}} \right) &= \int\limits_{{s_{\min }}}^{{s_0}} {\int\limits_{{u_{\min }}}^{{u_0}} {dsdu} } \left( {\begin{array}{*{20}{c}}
			{\frac{{{m_c}({m_c} + {m_b})}}{{m_{{B_c}}^2m_{{\eta _c}}^2{f_{{B_c}}}{f_{{\eta _c}}}}}}&{\frac{{{m_c}({m_c} + {m_b})}}{{m_{{B_c}}^2m_{{\eta _c}}^2{f_{{B_c}}}{f_{{\eta _c}}}}}}\\
			{\frac{{{Q^2}}}{{m_{{\eta _c}}^2 - m_{{B_c}}^2}} + \frac{1}{2}}&{\frac{{{Q^2}}}{{m_{{B_c}}^2 - m_{{\eta _c}}^2}} + \frac{1}{2}}
	\end{array}} \right)\left( {\begin{array}{*{20}{c}}
			{\rho _1^{\mathrm{QCD2}}(s,u,{Q^2})}\\
			{\rho _2^{\mathrm{QCD2}}(s,u,{Q^2})}
	\end{array}} \right)\\
	&\quad \times \exp ( - \frac{s}{{{M^2}}} - \frac{u}{{k{M^2}}})\exp (\frac{{m_{{B_c}}^2}}{{{M^2}}} + \frac{{m_{{\eta _c}}^2}}{{k{M^2}}})
\end{align}
\begin{align}
	{f_T}({Q^2}) =  - \frac{{{m_c}({m_b} + {m_c})({m_{{B_c}}} + {m_{{\eta _c}}})}}{{{f_{{\eta _c}}}m_{{\eta _c}}^2{f_{{B_c}}}m_{{B_c}}^2}}\int\limits_{{s_{\min }}}^{{s_0}} {\int\limits_{{u_{\min }}}^{{u_0}} {dsdu{\rho ^{\mathrm{QCD3}}}(s,u,{Q^2})\exp ( - \frac{s}{{{M^2}}} - \frac{u}{{k{M^2}}})} } \exp (\frac{{m_{{B_c}}^2}}{{{M^2}}} + \frac{{m_{{\eta _c}}^2}}{{k{M^2}}})
\end{align}
\begin{align}
	&{V}({Q^2}) = \int\limits_{{s_{min}}}^{{s_0}} {\int\limits_{{u_{min}}}^{{u_0}} {dsdu}{{{\rho }^{\mathrm{QCD4}}}(s,u,{Q^2})\exp ( - \frac{s}{{{M^2}}} - \frac{u}{{k{M^2}}})} } \frac{{({m_{{B_c}}} + {m_{J/\psi }})({m_c} + {m_b})}}{{{\rm{2}}{f_{{B_c}}}m_{{B_c}}^2{f_{J/\psi }}{m_{J/\psi }}}}\exp (\frac{{m_{{B_c}}^2}}{{{M^2}}}{\rm{ + }}\frac{{m_{J/\psi }^2}}{{k{M^2}}})
\end{align}
\begin{align}
	\notag
	\left( {\begin{array}{*{20}{c}}
			{{A_0}({Q^2})}\\
			{{A_1}({Q^2})}\\
			{{A_2}({Q^2})}
	\end{array}} \right)&=\int\limits_{{s_{\min }}}^{{s_0}} {\int\limits_{{u_{\min }}}^{{u_0}} {dsdu} } \left( {\begin{array}{*{20}{c}}
			{ - \frac{{{m_{J/\psi }}}}{{m_{{B_c}}^2 + m_{J/\psi }^2 + {Q^2}}}}&{\frac{{ - m_{{B_c}}^2 + m_{J/\psi }^2 + {Q^2}}}{{4{m_{J/\psi }}}}}&{\frac{{{m_{J/\psi }}(m_{{B_c}}^2 - m_{J/\psi }^2 + {Q^2})}}{{2(m_{{B_c}}^2 + m_{J/\psi }^2 + {Q^2})}}}\\
			{ - \frac{1}{{{m_{{B_c}}} + {m_{J/\psi }}}}}&0&0\\
			{ - \frac{{{m_{{B_c}}} + {m_{J/\psi }}}}{{m_{{B_c}}^2 + m_{J/\psi }^2 + {Q^2}}}}&{\frac{{{m_{{B_c}}} + {m_{J/\psi }}}}{2}}&{ - \frac{{m_{J/\psi }^2({m_{{B_c}}} + {m_{J/\psi }})}}{{m_{{B_c}}^2 + m_{J/\psi }^2 + {Q^2}}}}
	\end{array}} \right)\left( {\begin{array}{*{20}{c}}
			{\rho _1^{\mathrm{QCD5}}(s,u,{Q^2})}\\
			{\rho _2^{\mathrm{QCD5}}(s,u,{Q^2})}\\
			{\rho _3^{\mathrm{QCD5}}(s,u,{Q^2})}
	\end{array}} \right)\\
	&\quad \times \frac{{{m_b} + {m_c}}}{{{m_{J/\psi }}m_{{B_c}}^2{f_{J/\psi }}{f_{{B_c}}}}}\exp ( - \frac{s}{{{M^2}}} - \frac{u}{{k{M^2}}})\exp (\frac{{m_{{B_c}}^2}}{{{M^2}}} + \frac{{m_{{\eta _c}}^2}}{{k{M^2}}})
\end{align}
\begin{align}
	\notag
	{T_0}({Q^2}) =  - {({m_{{B_c}}} + {m_{J/\psi }})^2}\int\limits_{{s_{\min }}}^{{s_0}} {\int\limits_{{u_{\min }}}^{{u_0}} {dsdu\rho _1^{\mathrm{QCD6}}(s,u,{Q^2})\exp ( - \frac{s}{{{M^2}}} - \frac{u}{{k{M^2}}})} } \exp (\frac{{m_{{B_c}}^2}}{{{M^2}}} + \frac{{m_{{\eta _c}}^2}}{{k{M^2}}})\\
	\notag
	{T_1}({Q^2}) = \int\limits_{{s_{\min }}}^{{s_0}} {\int\limits_{{u_{\min }}}^{{u_0}} {dsdu\rho _2^{\mathrm{QCD6}}(s,u,{Q^2})\exp ( - \frac{s}{{{M^2}}} - \frac{u}{{k{M^2}}})} } \exp (\frac{{m_{{B_c}}^2}}{{{M^2}}} + \frac{{m_{{\eta _c}}^2}}{{k{M^2}}})\\
	{T_2}({Q^2}) = \int\limits_{{s_{\min }}}^{{s_0}} {\int\limits_{{u_{\min }}}^{{u_0}} {dsdu\rho _3^{\mathrm{QCD6}}(s,u,{Q^2})\exp ( - \frac{s}{{{M^2}}} - \frac{u}{{k{M^2}}})} } \exp (\frac{{m_{{B_c}}^2}}{{{M^2}}} + \frac{{m_{{\eta _c}}^2}}{{k{M^2}}})
\end{align}

\newpage
\section{The graphs of pole contributions and the Borel platform}\label{A2}
	
	\begin{figure*}[htbp]
		\centering
		\includegraphics[width=18cm]{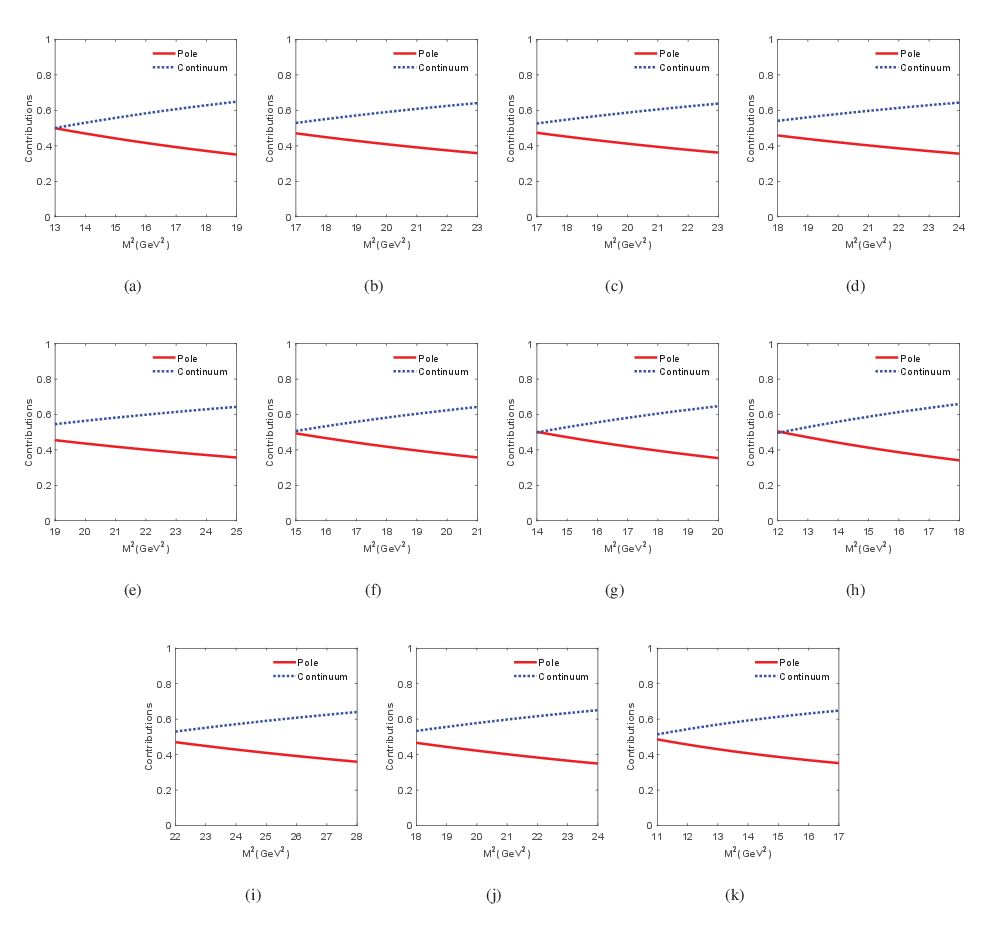}
		\caption{The pole and continuum contributions with variation of the Borel parameters. (a-d) are for the form factors $f_S, f_+, f_0$ and $f_T$ of decay process $B_c \to \eta_c$, (e-k) are for $V, A_0, A_1, A_2, T_0, T_1$ and $T_2$ of decay process $B_c \to J/\psi$.}
		\label{PC}
	\end{figure*}

	\begin{figure*}[htbp]
		\centering
		\includegraphics[width=18cm]{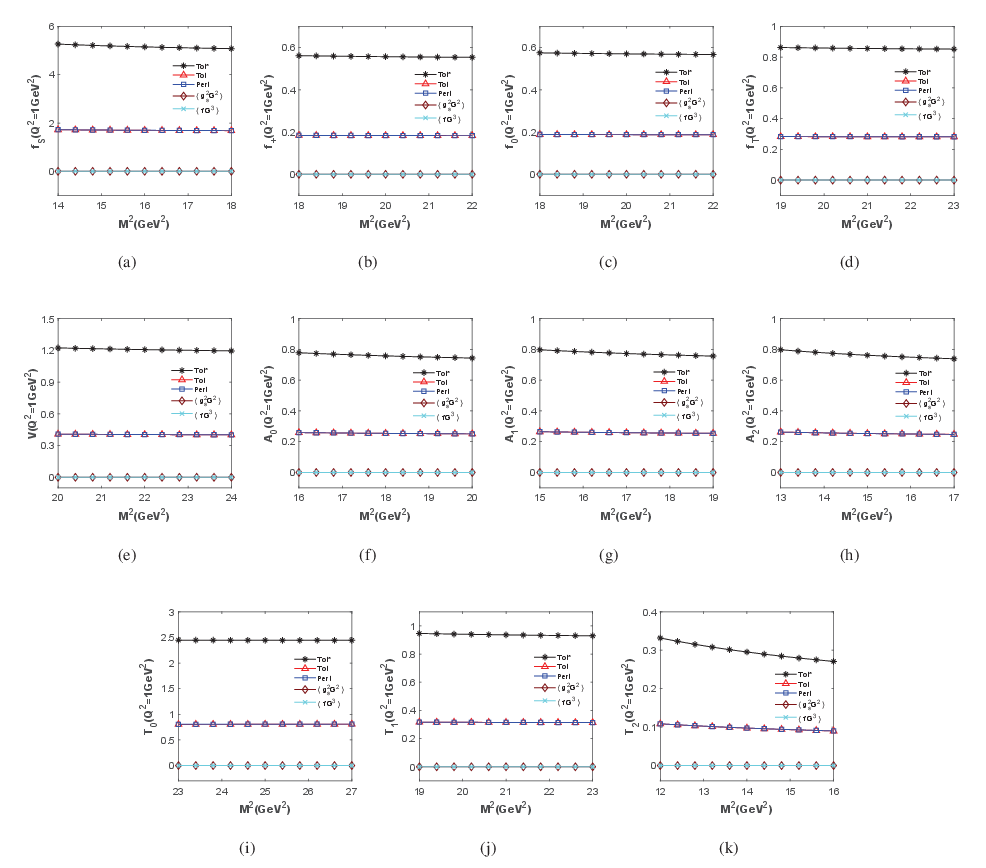}
		\caption{The contributions of perturbative part and different vacuum condensate terms with variation of the Borel parameter. The legend with star denotes the results obtained with Coulomb-like correction.}
		\label{BW}
	\end{figure*}
\FloatBarrier

\clearpage
\section{The decay widths of nonleptonic decay}\label{A3}
		\begin{align}
			\Gamma_{B^-_c \to \eta_c \pi^-}= \frac{{\sqrt {\lambda (m_{{B_c}}^2,m_{{\eta _c}}^2,m_\pi ^2)} }}{{{\rm{32}}\pi m_{{B_c}}^{\rm{3}}}}{(m_{{B_c}}^2 - m_{{\eta _c}}^2)^2}G_F^2|{V_{cb}}{|^2}|{V_{ud}}{|^2}{a_1^2}f_\pi ^2f_0^2({q^2})\left| {_{{q^2} = m_\pi ^2}} \right.
		\end{align}
		\begin{align}
			\Gamma_{B^-_c \to \eta_c K^-} = \frac{{\sqrt {\lambda (m_{{B_c}}^2,m_{{\eta _c}}^2,m_K^2)} }}{{32\pi m_{{B_c}}^3}}{(m_{{B_c}}^2 - m_{{\eta _c}}^2)^2}G_F^2|{V_{cb}}{|^2}|{V_{us}}{|^2}{a_1^2}f_K^2{f_0}{({q^2})^2}\left| {_{{q^2} = m_K^2}} \right.
		\end{align}
		\begin{align}
			\Gamma_{B^-_c \to \eta_c \rho^{-}} = \frac{{\sqrt {\lambda (m_{{B_c}}^2,m_{{\eta _c}}^2,m_{\rho}^2)} }}{{32\pi m_{{B_c}}^3}}[m_{{B_c}}^2 - {({m_{{\eta _c}}} + {m_{\rho}})^2}][m_{{B_c}}^2 - {({m_{{\eta _c}}} - {m_{\rho}})^2}]G_F^2|{V_{cb}}{|^2}|{V_{ud}}{|^2}{a_1^2}f_{\rho}^2{f_ + }{({q^2})^2}\left| {_{{q^2} = m_{\rho}^2}} \right.
		\end{align}
		\begin{align}
			\Gamma_{B^-_c \to \eta_c K^{*-}} = \frac{{\sqrt {\lambda (m_{{B_c}}^2,m_{{\eta _c}}^2,m_{K^*}^2)} }}{{32\pi m_{{B_c}}^3}}[m_{{B_c}}^2 - {({m_{{\eta _c}}} + {m_{K^*}})^2}][m_{{B_c}}^2 - {({m_{{\eta _c}}} - {m_{K^*}})^2}]G_F^2|{V_{cb}}{|^2}|{V_{us}}{|^2}{a_1^2}f_{K^*}^2{f_ + }{({q^2})^2}\left| {_{{q^2} = m_{K^*}^2}} \right.
		\end{align}
		\begin{align}
			\Gamma_{B^-_c \to J/\psi \pi^-}  = \frac{{\sqrt {\lambda (m_{{B_c}}^2,m_{J/\psi }^2,m_\pi ^2)} }}{{8\pi m_{{B_c}}^3}}[\frac{{{{(m_{{B_c}}^2 - m_{J/\psi }^2 - m_\pi ^2)}^2}}}{{4m_{J/\psi }^2}} - m_\pi ^2]m_{J/\psi }^2G_F^2|{V_{cb}}{|^2}|{V_{ud}}{|^2}{a_1^2}f_\pi ^2{A_0}{({q^2})^2}\left| {_{{q^2} = m_\pi ^2}} \right.
		\end{align}
		\begin{align}
			\Gamma_{B^-_c \to J/\psi K^-} = \frac{{\sqrt {\lambda (m_{{B_c}}^2,m_{J/\psi }^2,m_K^2)} }}{{8\pi m_{{B_c}}^3}}[\frac{{{{(m_{{B_c}}^2 - m_{J/\psi }^2 - m_K^2)}^2}}}{{4m_{J/\psi }^2}} - m_K^2]m_{J/\psi }^2G_F^2|{V_{cb}}{|^2}|{V_{us}}{|^2}{a_1^2}f_K^2{A_0}{({q^2})^2}\left| {_{{q^2} = m_K^2}} \right.
		\end{align}
		\begin{align}
			\notag
			\Gamma_{B^-_c \to J/\psi \rho^{-}}  =& \frac{{\sqrt {\lambda (m_{{B_c}}^2,m_{J/\psi }^2,m_{\rho}^2)} }}{{16\pi m_{{B_c}}^2}}G_F^2|{V_{cb}}{|^2}|{V_{cs}}{|^2}{a_1^2}f_{\rho}^2m_{\rho}^2\\
			\notag
			&\times \{  - \{ \frac{{8V{{({q^2})}^2}}}{{{{({m_{{B_c}}} + {m_{J/\psi }})}^2}}}[m_{{B_c}}^2m_{J/\psi }^2 - {(\frac{{m_{{B_c}}^2 + m_{J/\psi }^2 - m_{\rho}^2}}{2})^2}]\\
			\notag
			&- 3{({m_{{B_c}}} + {m_{J/\psi }})^2}{A_1}{({q^2})^2}\\
			\notag
			&- 2({m_{{B_c}}} + {m_{J/\psi }}){A_1}({q^2})B({q^2})[ - \frac{{m_{{B_c}}^2 - m_{J/\psi }^2 + m_{\rho}^2}}{2} + \frac{{(m_{{B_c}}^2 - m_{J/\psi }^2 - m_{\rho}^2)(m_{{B_c}}^2 + m_{J/\psi }^2 - m_{\rho}^2)}}{{4m_{J/\psi }^2}}]\\
			\notag
			&+ [ - m_{\rho}^2 + \frac{{{{(m_{{B_c}}^2 - m_{J/\psi }^2 - m_{\rho}^2)}^2}}}{{4m_{J/\psi }^2}}][B({q^2})m_{{B_c}}^2 - B({q^2})C({q^2})(m_{{B_c}}^2 + m_{J/\psi }^2 - m_{\rho}^2) + C{({q^2})^2}m_{J/\psi }^2]\} \\
			&+\frac{1}{{m_{\rho}^2}}{A_0}{({q^2})^2}[{(m_{{B_c}}^2 - m_{J/\psi }^2 - m_{\rho}^2)^2} - 4m_{\rho}^2m_{J/\psi }^2] \}
		\end{align}
		\begin{align}
			\notag
			\Gamma_{B^-_c \to J/\psi K^{*-}}  =& \frac{{\sqrt {\lambda (m_{{B_c}}^2,m_{J/\psi }^2,m_{K^*}^2)} }}{{16\pi m_{{B_c}}^2}}G_F^2|{V_{cb}}{|^2}|{V_{ud}}{|^2}{a_1^2}f_{K^*}^2m_{K^*}^2\\
			\notag
			&\times \{  - \{ \frac{{8V{{({q^2})}^2}}}{{{{({m_{{B_c}}} + {m_{J/\psi }})}^2}}}[m_{{B_c}}^2m_{J/\psi }^2 - {(\frac{{m_{{B_c}}^2 + m_{J/\psi }^2 - m_{K^*}^2}}{2})^2}]\\
			\notag
			&- 3{({m_{{B_c}}} + {m_{J/\psi }})^2}{A_1}{({q^2})^2}\\
			\notag
			&- 2({m_{{B_c}}} + {m_{J/\psi }}){A_1}({q^2})B({q^2})[ - \frac{{m_{{B_c}}^2 - m_{J/\psi }^2 + m_{K^*}^2}}{2} + \frac{{(m_{{B_c}}^2 - m_{J/\psi }^2 - m_{K^*}^2)(m_{{B_c}}^2 + m_{J/\psi }^2 - m_{K^*}^2)}}{{4m_{J/\psi }^2}}]\\
			\notag
			&+ [ - m_{K^*}^2 + \frac{{{{(m_{{B_c}}^2 - m_{J/\psi }^2 - m_{K^*}^2)}^2}}}{{4m_{J/\psi }^2}}][B({q^2})m_{{B_c}}^2 - B({q^2})C({q^2})(m_{{B_c}}^2 + m_{J/\psi }^2 - m_{K^*}^2) + C{({q^2})^2}m_{J/\psi }^2]\} \\
			&+\frac{1}{{m_{K^*}^2}}{A_0}{({q^2})^2}[{(m_{{B_c}}^2 - m_{J/\psi }^2 - m_{K^*}^2)^2} - 4m_{K^*}^2m_{J/\psi }^2] \}
		\end{align}
\begin{flalign}
	\notag
	&\mathrm{where}&
\end{flalign}
	\begin{align}
		\notag
		&B({q^2}) = \frac{{{m_{{B_c}}} + {m_{J/\psi }}}}{{{q^2}}}{A_1}({q^2}) + \frac{{{A_2}({q^2})}}{{{m_{{B_c}}} + {m_{J/\psi }}}} - \frac{{({m_{{B_c}}} - {m_{J/\psi }}){A_2}({q^2})}}{{{q^2}}} - \frac{{2{m_{J/\psi }}{A_0}({q^2})}}{{{q^2}}}\\
		&C({q^2}) = \frac{{{m_{{B_c}}} + {m_{J/\psi }}}}{{{q^2}}}{A_1}({q^2}) - \frac{{{A_2}({q^2})}}{{{m_{{B_c}}} + {m_{J/\psi }}}} - \frac{{({m_{{B_c}}} - {m_{J/\psi }}){A_2}({q^2})}}{{{q^2}}} - \frac{{2{m_{J/\psi }}{A_0}({q^2})}}{{{q^2}}}
	\end{align}
\section{The differential decay widths of semileptonic decay}\label{A4}		
		\begin{align}
			\notag
			\frac{{d\Gamma_{B^-_c \to \eta_c \mathrm{l} \bar\nu } }}{{d{q^2}}} =& - \frac{{G_F^2|{V_{cb}}{|^2}\sqrt {\lambda (m_{{B_c}}^2,m_{{\eta _c}}^2,{q^2})} \sqrt {\lambda ({q^2},m_l^2,0)} }}{{384{\pi ^3}m_{{B_c}}^3{{({q^2})}^4}}}({q^2} - m_l^2)\\
			\notag
			&\times \{ (2{q^2} + m_l^2)[{D(q^2)^2}m_{{B_c}}^2 + D(q^2)E(q^2)(m_{{B_c}}^2 - {q^2} + m_{{\eta _c}}^2) + {E(q^2)^2}m_{{\eta _c}}^2] \\
			&- 2{q^2}({q^2} + 2m_l^2){f_0}{({q^2})^2}{(m_{{B_c}}^2 - m_{{\eta _c}}^2)^2}\}
		\end{align}
		
		\begin{align}
			\notag
			\frac{{d\Gamma_{B^-_c \to J/\psi \mathrm{l} \bar\nu} }}{{d{q^2}}} = &  - \frac{{G_F^2|{V_{cb}}{|^2}\sqrt {\lambda (m_{{B_c}}^2,m_{J/\psi }^2,{q^2})} \sqrt {\lambda ({q^2},m_l^2,0)} }}{{192{\pi ^3}m_{{B_c}}^3{{({q^2})}^2}}}({q^2} - m_l^2)\\
			\notag
			&\times \{ (2{q^2} + m_l^2)\{ \frac{{8V{{({q^2})}^2}}}{{{{({m_{{B_c}}} + {m_{J/\psi }})}^2}}}[m_{{B_c}}^2m_{J/\psi }^2 - {(\frac{{m_{{B_c}}^2 + m_{J/\psi }^2 - {q^2}}}{2})^2}]\\
			\notag
			&- 3{({m_{{B_c}}} + {m_{J/\psi }})^2}{A_1}{({q^2})^2}\\
			\notag
			&- 2({m_{{B_c}}} + {m_{J/\psi }}){A_1}({q^2})B({q^2})[ - \frac{{m_{{B_c}}^2 - m_{J/\psi }^2 + {q^2}}}{2} + \frac{{(m_{{B_c}}^2 - m_{J/\psi }^2 - {q^2})(m_{{B_c}}^2 + m_{J/\psi }^2 - {q^2})}}{{4m_{J/\psi }^2}}]\\
			\notag
			&+ [ - {q^2} + \frac{{{{(m_{{B_c}}^2 - m_{J/\psi }^2 -{q^2})}^2}}}{{4m_{J/\psi }^2}}][B({q^2})m_{{B_c}}^2 - B({q^2})C({q^2})(m_{{B_c}}^2 + m_{J/\psi }^2 - {q^2}) + C{({q^2})^2}m_{J/\psi }^2]\} \\
			& - \frac{{({q^2} + 2m_l^2)}}{{2{q^2}}}{A_0}{({q^2})^2}[{(m_{{B_c}}^2 - m_{J/\psi }^2 - {q^2})^2} - 4{q^2}m_{J/\psi }^2]
		\end{align}
		\begin{flalign}
			\notag
			&\mathrm{where}&
		\end{flalign}
		\begin{align}
			\notag
			&B({q^2}) = \frac{{{m_{{B_c}}} + {m_{J/\psi }}}}{{{q^2}}}{A_1}({q^2}) + \frac{{{A_2}({q^2})}}{{{m_{{B_c}}} + {m_{J/\psi }}}} - \frac{{({m_{{B_c}}} - {m_{J/\psi }}){A_2}({q^2})}}{{{q^2}}} - \frac{{2{m_{J/\psi }}{A_0}({q^2})}}{{{q^2}}}\\
			\notag
			&C({q^2}) = \frac{{{m_{{B_c}}} + {m_{J/\psi }}}}{{{q^2}}}{A_1}({q^2}) - \frac{{{A_2}({q^2})}}{{{m_{{B_c}}} + {m_{J/\psi }}}} - \frac{{({m_{{B_c}}} - {m_{J/\psi }}){A_2}({q^2})}}{{{q^2}}} - \frac{{2{m_{J/\psi }}{A_0}({q^2})}}{{{q^2}}}\\
			\notag
			&D({q^2}) = f_ + ^{{B_c} \to {\eta _c}}({q^2}){q^2} + (f_0^{{B_c} \to {\eta _c}}({q^2}) - f_ + ^{{B_c} \to {\eta _c}}({q^2}))(m_{{B_c}}^2 - m_{{\eta _c}}^2)\\
			&E({q^2}) = f_ + ^{{B_c} \to {\eta _c}}({q^2}){q^2} - (f_0^{{B_c} \to {\eta _c}}({q^2}) - f_ + ^{{B_c} \to {\eta _c}}({q^2}))(m_{{B_c}}^2 - m_{{\eta _c}}^2)
		\end{align}
	\end{widetext}



\begin{thebibliography}{}
	
\bibitem{CDF:1998axz}
F.~Abe \textit{et al.} [CDF],
Observation of $B_c$ mesons in $p\bar{p}$ collisions at $\sqrt{s} = 1.8$ TeV,
\href{https://doi.org/10.1103/PhysRevD.58.112004}{Phys. Rev. D \textbf{58}, 112004 (1998)}.

\bibitem{LHCb:2017vlu}
R.~Aaij \textit{et al.} [LHCb],
Measurement of the ratio of branching fractions $\mathcal{B}(B_c^+\,\to\,J/\psi\tau^+\nu_\tau)$/$\mathcal{B}(B_c^+\,\to\,J/\psi\mu^+\nu_\mu)$,
\href{https://doi.org/10.1103/PhysRevLett.120.121801}{Phys. Rev. Lett. \textbf{120}, no.12, 121801 (2018)}.

\bibitem{LHCb:2012ihf}
R.~Aaij \textit{et al.} [LHCb],
Measurements of $B_c^+$ production and mass with the $B_c^+ \to J/\psi \pi^+$ decay,
\href{https://doi.org/10.1103/PhysRevLett.109.232001}{Phys. Rev. Lett. \textbf{109}, 232001 (2012)}.

\bibitem{LHCb:2012ag}
R.~Aaij \textit{et al.} [LHCb],
First observation of the decay $B_c^+ \to J/\psi \pi^+\pi^-\pi^+$,
\href{https://doi.org/10.1103/PhysRevLett.108.251802}{Phys. Rev. Lett. \textbf{108}, 251802 (2012)}.

\bibitem{LHCb:2013hwj}
R.~Aaij \textit{et al.} [LHCb],
First observation of the decay $B_{c}^{+}\to J/\psi K^+$,
\href{https://doi.org/10.1007/JHEP09(2013)075}{JHEP \textbf{09}, 075 (2013)}.

\bibitem{LHCb:2013kwl}
R.~Aaij \textit{et al.} [LHCb],
Observation of $B^+_c \to J/\psi D_s^+$ and $B^+_c \to J/\psi D_s^{*+}$ decays,
\href{https://doi.org/10.1103/PhysRevD.87.112012}{Phys. Rev. D \textbf{87}, no.11, 112012 (2013)}.

\bibitem{LHCb:2013rud}
R.~Aaij \textit{et al.} [LHCb],
Observation of the decay $B_c \to J/\psi K^+ K^- \pi^+ $,
\href{https://doi.org/10.1007/JHEP11(2013)094}{JHEP \textbf{11}, 094 (2013)}.

\bibitem{Gao:2010zzc}
Y.~N.~Gao, J.~He, P.~Robbe, M.~H.~Schune and Z.~W.~Yang,
Experimental prospects of the B(c) studies of the LHCb experiment,
\href{https://doi.org/10.1088/0256-307X/27/6/061302}{Chin. Phys. Lett. \textbf{27}, 061302 (2010)}.

\bibitem{Colquhoun:2016osw}
B.~Colquhoun \textit{et al.} [HPQCD],
$B_c$ decays from highly improved staggered quarks and NRQCD,
\href{https://doi.org/10.22323/1.256.0281}{PoS \textbf{LATTICE2016}, 281 (2016)}.

\bibitem{Harrison:2020gvo}
J.~Harrison \textit{et al.} [HPQCD],
$B_c \to J/\psi$ form factors for the full $q^2$ range from lattice QCD,
\href{https://doi.org/10.1103/PhysRevD.102.094518}{Phys. Rev. D \textbf{102}, no.9, 094518 (2020)}.

\bibitem{Nobes:2000pm}
M.~A.~Nobes and R.~M.~Woloshyn,
Decays of the $B_c$ meson in a relativistic quark meson model,
\href{https://doi.org/10.1088/0954-3899/26/7/308}{J. Phys. G \textbf{26}, 1079-1094 (2000)}.

\bibitem{Faustov:2022ybm}
R.~N.~Faustov, V.~O.~Galkin and X.~W.~Kang,
Relativistic description of the semileptonic decays of bottom mesons,
\href{https://doi.org/10.1103/PhysRevD.106.013004}{Phys. Rev. D \textbf{106}, no.1, 013004 (2022)}.

\bibitem{Tran:2018kuv}
C.~T.~Tran, M.~A.~Ivanov, J.~G.~K\"orner and P.~Santorelli,
Implications of new physics in the decays $B_c \to (J/\psi,\eta_c)\tau\nu$,
\href{https://doi.org/10.1103/PhysRevD.97.054014}{Phys. Rev. D \textbf{97}, no.5, 054014 (2018)}.

\bibitem{Qiao:2012hp}
C.~F.~Qiao, P.~Sun, D.~Yang and R.~L.~Zhu,
B$_c$ exclusive decays to charmonium and a light meson at next-to-leading order accuracy,
\href{https://doi.org/10.1103/PhysRevD.89.034008}{Phys. Rev. D \textbf{89}, no.3, 034008 (2014)}.

\bibitem{Tang:2022nqm}
R.~Y.~Tang, Z.~R.~Huang, C.~D.~L\"u and R.~Zhu,
Scrutinizing new physics in semileptonic $B_{c} \to J/\psi {\tau}{\nu}$ decay,
\href{https://doi.org/10.1088/1361-6471/ac8d1e}{J. Phys. G \textbf{49} (2022) no.11, 115003}.

\bibitem{Zhang:2023ypl}
Z.~Q.~Zhang, Z.~J.~Sun, Y.~C.~Zhao, Y.~Y.~Yang and Z.~Y.~Zhang,
Covariant light-front approach for $B_c$ decays into charmonium: implications on form factors and branching ratios,
\href{https://doi.org/10.1140/epjc/s10052-023-11576-x}{Eur. Phys. J. C \textbf{83}, no.6, 477 (2023)}.

\bibitem{Shi:2016gqt}
Y.~J.~Shi, W.~Wang and Z.~X.~Zhao,
$B_c\rightarrow B_{sJ}$ form factors and $B_c$ decays into $B_{sJ}$ in covariant light-front approach,
\href{https://doi.org/10.1140/epjc/s10052-016-4405-1}{Eur. Phys. J. C \textbf{76}, no.10, 555 (2016)}.

\bibitem{Azizi:2013zta}
K.~Azizi, Y.~Sarac and H.~Sundu,
Investigation of the $B_c\to \chi_{c2} l \bar{\nu} $ transition via QCD sum rules,
\href{https://doi.org/10.1140/epjc/s10052-013-2638-9}{Eur. Phys. J. C \textbf{73}, no.11, 2638 (2013)}.

\bibitem{Kiselev:2000pp}
V.~V.~Kiselev, A.~E.~Kovalsky and A.~K.~Likhoded,
$B_c$ decays and lifetime in QCD sum rules,
\href{https://doi.org/10.1016/S0550-3213(00)00386-2}{Nucl. Phys. B \textbf{585}, 353-382 (2000)}.

\bibitem{Azizi:2019aaf}
K.~Azizi, Y.~Sarac and H.~Sundu,
Lepton flavor universality violation in semileptonic tree level weak transitions,
\href{https://doi.org/10.1103/PhysRevD.99.113004}{Phys. Rev. D \textbf{99}, no.11, 113004 (2019)}.

\bibitem{Colangelo:1992cx}
P.~Colangelo, G.~Nardulli and N.~Paver,
QCD sum rules calculation of B(c) decays,
\href{https://doi.org/10.1007/BF01555737}{Z. Phys. C \textbf{57}, 43-50 (1993)}.

\bibitem{Wang:2007fs}
Y.~M.~Wang and C.~D.~Lu,
Weak productions of new charmonium in semileptonic decays of $B_c$,
\href{https://doi.org/10.1103/PhysRevD.77.054003}{Phys. Rev. D \textbf{77}, 054003 (2008)}.

\bibitem{Huang:2007kb}
T.~Huang and F.~Zuo,
Semileptonic $B_c$ decays and charmonium distribution amplitude,
\href{https://doi.org/10.1140/epjc/s10052-007-0333-4}{Eur. Phys. J. C \textbf{51}, 833-839 (2007)}.

\bibitem{Leljak:2019eyw}
D.~Leljak, B.~Melic and M.~Patra,
On lepton flavour universality in semileptonic $B_{c} \to \eta_{c}, J/{\psi}$ decays,
\href{https://doi.org/10.1007/JHEP05(2019)094}{JHEP \textbf{05}, 094 (2019)}.

\bibitem{Bordone:2022drp}
M.~Bordone, A.~Khodjamirian and T.~Mannel,
New sum rules for the $B_{c}\to J/{\psi}$ form factors,
\href{https://doi.org/10.1007/JHEP01(2023)032}{JHEP \textbf{01}, 032 (2023)}.

\bibitem{Wang:2012lrc}
W.~F.~Wang, Y.~Y.~Fan and Z.~J.~Xiao,
Semileptonic decays $B_c\to(\eta_c,J/\psi)l\nu$ in the perturbative QCD approach,
\href{https://doi.org/10.1088/1674-1137/37/9/093102}{Chin. Phys. C \textbf{37}, 093102 (2013)}.

\bibitem{Liu:2023kxr}
X.~Liu,
Bc-meson decays into J/\ensuremath{\psi} plus a light meson in the improved perturbative QCD formalism,
\href{https://doi.org/10.1103/PhysRevD.108.096006}{Phys. Rev. D \textbf{108}, no.9, 096006 (2023)}.

\bibitem{Liu:2020upy}
X.~Liu, H.~n.~Li and Z.~J.~Xiao,
Next-to-leading-logarithm $k_T$ resummation for $B_c\to J/\psi$ decays,
\href{https://doi.org/10.1016/j.physletb.2020.135892}{Phys. Lett. B \textbf{811}, 135892 (2020)}.

\bibitem{Liu:2018kuo}
X.~Liu, H.~n.~Li and Z.~J.~Xiao,
Improved perturbative QCD formalism for $B_c$ meson decays,
\href{https://doi.org/10.1103/PhysRevD.97.113001}{Phys. Rev. D \textbf{97}, no.11, 113001 (2018)}.

\bibitem{Shifman:1978bx}
M.~A.~Shifman, A.~I.~Vainshtein and V.~I.~Zakharov,
QCD and Resonance Physics. Theoretical Foundations,
\href{https://doi.org/10.1016/0550-3213(79)90022-1}{Nucl. Phys. B \textbf{147}, 385-447 (1979)}.

\bibitem{Shifman:1978by}
M.~A.~Shifman, A.~I.~Vainshtein and V.~I.~Zakharov,
QCD and Resonance Physics: Applications,
\href{https://doi.org/10.1016/0550-3213(79)90023-3}{Nucl. Phys. B \textbf{147}, 448-518 (1979)}.

\bibitem{Wei:2006wa}
W.~Wei, X.~Liu and S.~L.~Zhu,
D wave heavy mesons,
\href{https://doi.org/10.1103/PhysRevD.75.014013}{Phys. Rev. D \textbf{75}, 014013 (2007)}.

\bibitem{Wang:2012kw}
Z.~G.~Wang,
Analysis of the vector and axialvector $B_c$ mesons with QCD sum rules,
\href{https://doi.org/10.1140/epja/i2013-13131-7}{Eur. Phys. J. A \textbf{49}, 131 (2013)}.

\bibitem{Aliev:2012iv}
T.~M.~Aliev, K.~Azizi and M.~Savci,
The masses and residues of doubly heavy spin-3/2 baryons,
\href{https://doi.org/10.1088/0954-3899/40/6/065003}{J. Phys. G \textbf{40}, 065003 (2013)}.

\bibitem{Aliev:2012tt}
T.~M.~Aliev, K.~Azizi and M.~Savci,
Masses and Residues of the Triply Heavy Spin-1/2 Baryons,
\href{https://doi.org/10.1007/JHEP04(2013)042}{JHEP \textbf{04}, 042 (2013)}.

\bibitem{Belyaev:1993wp}
V.~M.~Belyaev, A.~Khodjamirian and R.~Ruckl,
QCD calculation of the $B \to \pi, K$ form-factors,
\href{https://doi.org/10.1007/BF01474633}{Z. Phys. C \textbf{60}, 349-356 (1993)}.

\bibitem{Dai:1996xv}
Y.~B.~Dai, C.~S.~Huang, M.~Q.~Huang and C.~Liu,
QCD sum rule analysis for the $\Lambda(b) \to \Lambda(c)$ semileptonic decay,
\href{https://doi.org/10.1016/0370-2693(96)01029-5}{Phys. Lett. B \textbf{387}, 379-385 (1996)}.

\bibitem{Yang:2005bv}
M.~Z.~Yang,
Semileptonic decay of $B$ and $D \to K^*_0 (1430) \bar l \nu$ from QCD sum rule,
\href{https://doi.org/10.1103/PhysRevD.73.079901}{Phys. Rev. D \textbf{73}, 034027 (2006)}.
[erratum: Phys. Rev. D \textbf{73}, 079901 (2006)]

\bibitem{Shi:2019hbf}
Y.~J.~Shi, W.~Wang and Z.~X.~Zhao,
QCD Sum Rules Analysis of Weak Decays of Doubly-Heavy Baryons,
\href{https://doi.org/10.1140/epjc/s10052-020-8096-2}{Eur. Phys. J. C \textbf{80}, no.6, 568 (2020)}.

\bibitem{Zhao:2020mod}
Z.~X.~Zhao, R.~H.~Li, Y.~L.~Shen, Y.~J.~Shi and Y.~S.~Yang,
The semileptonic form factors of $\Lambda_{b}\to\Lambda_{c}$ and $\Xi_{b}\to\Xi_{c}$ in QCD sum rules,
\href{https://doi.org/10.1140/epjc/s10052-020-08767-1}{Eur. Phys. J. C \textbf{80}, no.12, 1181 (2020)}.

\bibitem{Zhang:2023nxl}
S.~Q.~Zhang and C.~F.~Qiao,
\ensuremath{\Lambda}c semileptonic decays,
\href{https://doi.org/10.1103/PhysRevD.108.074017}{Phys. Rev. D \textbf{108}, no.7, 074017 (2023)}.

\bibitem{Lu:2024tgy}
J.~Lu, G.~L.~Yu, Z.~G.~Wang and B.~Wu,
Analysis of the electromagnetic form factors and the radiative decays of the vector heavy-light mesons,
\href{https://doi.org/10.1016/j.physletb.2024.138624}{Phys. Lett. B \textbf{852}, 138624 (2024)}.

\bibitem{Bracco:2011pg}
M.~E.~Bracco, M.~Chiapparini, F.~S.~Navarra and M.~Nielsen,
Charm couplings and form factors in QCD sum rules,
\href{https://doi.org/10.1016/j.ppnp.2012.03.002}{Prog. Part. Nucl. Phys. \textbf{67}, 1019-1052 (2012)}

\bibitem{Yu:2015xwa}
G.~L.~Yu, Z.~Y.~Li and Z.~G.~Wang,
Analysis of the strong coupling constant $G_{D_{s}^{*}D_{s}\phi}$ and the decay width of $D_{s}^{*}\to D_{s}\gamma$ with QCD sum rules,
\href{https://doi.org/10.1140/epjc/s10052-015-3460-3}{Eur. Phys. J. C \textbf{75}, no.6, 243 (2015)}.

\bibitem{Lu:2023pcg}
J.~Lu, G.~L.~Yu, Z.~G.~Wang and B.~Wu,
Analysis of the strong vertices of $\Sigma _{c}\Delta D^{*}$ and $\Sigma _{b}\Delta B^{*}$ in QCD sum rules,
\href{https://doi.org/10.1140/epjc/s10052-023-12076-8}{Eur. Phys. J. C \textbf{83} (2023) no.10, 907}.

\bibitem{Lu:2023gmd}
J.~Lu, G.~L.~Yu and Z.~G.~Wang,
The strong vertices of charmed mesons $D$, $D^{*}$ and charmonia $J/\psi $, $\eta _{c}$,
\href{https://doi.org/10.1140/epja/s10050-023-01115-3}{Eur. Phys. J. A \textbf{59} (2023) no.8, 195}.

\bibitem{Lu:2023lvu}
J.~Lu, G.~L.~Yu, Z.~G.~Wang and B.~Wu,
Strong vertices of bottom mesons $B$ and $B^*$ and bottomonia $\Upsilon$, ${\eta}_b *$,
\href{https://doi.org/10.1088/1674-1137/ad061d}{Chin. Phys. C \textbf{48} (2024) no.1, 013102}.

\bibitem{Colangelo:2000dp}
P.~Colangelo and A.~Khodjamirian,
QCD sum rules, a modern perspective,
\href{https://arxiv.org/abs/hep-ph/0010175}{arXiv:hep-ph/0010175 [hep-ph]}.

\bibitem{Colangelo:2022lpy}
P.~Colangelo, F.~De Fazio, F.~Loparco, N.~Losacco and M.~Novoa-Brunet,
Relations among $B_{c}\to J/{\psi}, {\eta}_{c}$ form factors,
\href{https://doi.org/10.1007/JHEP09(2022)028}{JHEP \textbf{09}, 028 (2022)}.

\bibitem{Reinders:1984sr}
L.~J.~Reinders, H.~Rubinstein and S.~Yazaki,
Hadron Properties from QCD Sum Rules,
\href{https://doi.org/10.1016/0370-1573(85)90065-1}{Phys. Rept. \textbf{127}, 1 (1985)}.

\bibitem{Cutkosky:1960sp}
R.~E.~Cutkosky,
Singularities and discontinuities of Feynman amplitudes,
\href{https://doi.org/10.1063/1.1703676}{J. Math. Phys. \textbf{1}, 429-433 (1960)}.

\bibitem{Kiselev:1993ea}
V.~V.~Kiselev and A.~V.~Tkabladze,
Semileptonic B(c) decays from QCD sum rules,
\href{https://doi.org/10.1103/PhysRevD.48.5208}{Phys. Rev. D \textbf{48}, 5208-5214 (1993)}.

\bibitem{Kiselev:1999sc}
V.~V.~Kiselev, A.~K.~Likhoded and A.~I.~Onishchenko,
Semileptonic $B_c$ meson decays in sum rules of QCD and NRQCD,
\href{https://doi.org/10.1016/S0550-3213(99)00505-2}{Nucl. Phys. B \textbf{569}, 473-504 (2000)}.

\bibitem{ParticleDataGroup:2022pth}
R.~L.~Workman \textit{et al.} [Particle Data Group],
Review of Particle Physics,
\href{https://doi.org/10.1093/ptep/ptac097}{PTEP \textbf{2022}, 083C01 (2022)}.

\bibitem{Wang:2013cha}
Z.~G.~Wang,
The radiative decays  $B_c^{*\pm} \to B_c^{\pm} \gamma $ with QCD sum rules,
\href{https://doi.org/10.1140/epjc/s10052-013-2559-7}{Eur. Phys. J. C \textbf{73}, no.9, 2559 (2013)}.

\bibitem{Wang:2013iia}
Z.~G.~Wang,
Analysis of hadronic coupling constants $G_{B_c^*B_c\Upsilon}$, $G_{B_c^*B_c J/\psi}$, $G_{B_cB_c\Upsilon}$ and $G_{B_cB_c J/\psi}$ with QCD sum rules,
\href{https://doi.org/10.1103/PhysRevD.89.034017}{Phys. Rev. D \textbf{89}, no.3, 034017 (2014)}


\bibitem{Narison:2010cg}
S.~Narison,
Gluon condensates and $c$, $b$ quark masses from quarkonia ratios of moments,
\href{https://doi.org/10.1016/j.physletb.2011.09.116}{Phys. Lett. B \textbf{693}, 559-566 (2010)}.
[erratum: Phys. Lett. B \textbf{705}, 544-544 (2011)]

\bibitem{Narison:2011xe}
S.~Narison,
Gluon Condensates and precise $\overline{m}_{c,b}$ from QCD-Moments and their ratios to Order $\alpha_s^3$ and $\left\langle {G^4} \right\rangle $,
\href{https://doi.org/10.1016/j.physletb.2011.11.058}{Phys. Lett. B \textbf{706}, 412-422 (2012)}.

\bibitem{Narison:2011rn}
S.~Narison,
Gluon Condensates and $\bar{m}_b(\bar{m}_b)$ from QCD-Exponential Moments at Higher Orders,
\href{https://doi.org/10.1016/j.physletb.2011.12.047}{Phys. Lett. B \textbf{707}, 259-263 (2012)}.


\bibitem{ZGW:2024Bc}
Z.~G.~Wang,
The $B_{c}$ meson and its scalar cousin with the QCD sum rules,
\href{https://doi.org/10.48550/arXiv.2401.12571}{arXiv:2401.12571 [hep-ph] (2024)}.

\bibitem{Narison:2020guz}
S.~Narison,
QCD parameters, $f_{B_c}$ and $f_{B_c(2S)}$ from relativistic heavy quark sum rules,
\href{https://doi.org/10.1016/j.nuclphysbps.2019.11.024}{Nucl. Part. Phys. Proc. \textbf{309-311}, 135-147 (2020)}.

\bibitem{Becirevic:2013bsa}
D.~Be\v{c}irevi\'c, G.~Duplan\v{c}i\'c, B.~Klajn, B.~Meli\'c and F.~Sanfilippo,
Lattice QCD and QCD sum rule determination of the decay constants of $\eta_c$, $J/\psi$ and $h_c$ states,
\href{https://doi.org/10.1016/j.nuclphysb.2014.03.024}{Nucl. Phys. B \textbf{883}, 306-327 (2014)}.


\bibitem{Wang:2007ys}
Y.~M.~Wang, H.~Zou, Z.~T.~Wei, X.~Q.~Li and C.~D.~Lu,
The Transition form-factors for semileptonic weak decays of $J/ \psi$ in QCD sum rules,
\href{https://doi.org/10.1140/epjc/s10052-007-0498-x}{Eur. Phys. J. C \textbf{54}, 107-121 (2008)}.

\bibitem{Wang:2008xt}
W.~Wang, Y.~L.~Shen and C.~D.~Lu,
Covariant Light-Front Approach for $B_c$ transition form factors,
\href{https://doi.org/10.1103/PhysRevD.79.054012}{Phys. Rev. D \textbf{79}, 054012 (2009)}.

\bibitem{Wang:2012vna}
Z.~G.~Wang,
The $B_c$-decays $B_c^+ \to J/\psi \pi^+\pi^-\pi^+$, $\eta_c \pi^+\pi^-\pi^+ $,
\href{https://doi.org/10.1103/PhysRevD.86.054010}{Phys. Rev. D \textbf{86}, 054010 (2012)}.

\bibitem{Biswas:2023bqz}
A.~Biswas, S.~Nandi and S.~Sahoo,
Analyzing the semileptonic and nonleptonic $B_c \to J/\psi, \eta_c$ decays,
\href{https://arxiv.org/abs/2311.00758}{[arXiv:2311.00758 [hep-ph]]}.

\bibitem{Boyd:1994tt}
C.~G.~Boyd, B.~Grinstein and R.~F.~Lebed,
Constraints on form-factors for exclusive semileptonic heavy to light meson decays,
\href{https://doi.org/10.1103/PhysRevLett.74.4603}{Phys. Rev. Lett. \textbf{74}, 4603-4606 (1995)}.

\bibitem{Wang:2015vgv}
Y.~M.~Wang and Y.~L.~Shen,
QCD corrections to $B \to {\pi}$ form factors from light-cone sum rules,
\href{https://doi.org/10.1016/j.nuclphysb.2015.07.016}{Nucl. Phys. B \textbf{898}, 563-604 (2015)}.

\bibitem{Bourrely:2008za}
C.~Bourrely, I.~Caprini and L.~Lellouch,
Model-independent description of $B \to \pi l \nu$ decays and a determination of $|V_{ub}|$,
\href{https://doi.org/10.1103/PhysRevD.82.099902}{Phys. Rev. D \textbf{79}, 013008 (2009)}.

\bibitem{Cui:2022zwm}
B.~Y.~Cui, Y.~K.~Huang, Y.~L.~Shen, C.~Wang and Y.~M.~Wang,
Precision calculations of $B_{d,s} \to {\pi}$, $K$ decay form factors in soft-collinear effective theory,
\href{https://doi.org/10.1007/JHEP03(2023)140}{JHEP \textbf{03}, 140 (2023)}.

\bibitem{Mohammadi:2014rpa}
B.~Mohammadi and H.~Mehraban,
Three-body decays of $B^{0(+)} \to K^{*0(+)} \pi^{+} \pi^{-}$,
\href{https://doi.org/10.1140/epja/i2014-14122-x}{Eur. Phys. J. A \textbf{50}, 122 (2014)}.

\bibitem{Mohammadi:2023itb}
B.~Mohammadi and S.~Khodadad,
Study of strange beauty neutral meson decays into vector and pseudoscalar mesons,
\href{https://doi.org/10.1007/s12648-023-02600-7}{Indian J. Phys. \textbf{97}, no.8, 2531-2536 (2023)}.


\bibitem{LHCb:2014ilr}
R.~Aaij \textit{et al.} [LHCb],
Measurement of the $B_c^+$ meson lifetime using $B_c^+ \to J\!/\!\psi \mu^+ \nu_{\mu} X$ decays,
\href{https://doi.org/10.1140/epjc/s10052-014-2839-x}{Eur. Phys. J. C \textbf{74}, no.5, 2839 (2014)}.

\bibitem{ParticleDataGroup:2020ssz}
P.~A.~Zyla \textit{et al.} [Particle Data Group],
\href{https://doi.org/10.1093/ptep/ptaa104}{PTEP \textbf{2020}, no.8, 083C01 (2020)}.

\bibitem{ParticleDataGroup:2010dbb}
K.~Nakamura \textit{et al.} [Particle Data Group],
Review of particle physics,
\href{https://doi.org/10.1088/0954-3899/37/7A/075021}{J. Phys. G \textbf{37}, 075021 (2010)}.

\bibitem{Buchalla:1995vs}
G.~Buchalla, A.~J.~Buras and M.~E.~Lautenbacher,
Weak decays beyond leading logarithms,
\href{https://doi.org/10.1103/RevModPhys.68.1125}{Rev. Mod. Phys. \textbf{68}, 1125-1144 (1996)}.

\bibitem{Kiselev:2002vz}
V.~V.~Kiselev,
Exclusive decays and lifetime of $B_c$ meson in QCD sum rules,
\href{https://arxiv.org/abs/hep-ph/0211021}{[arXiv:0211021 [hep-ph]]}.

\bibitem{Ebert:2003cn}
D.~Ebert, R.~N.~Faustov and V.~O.~Galkin,
Weak decays of the $B_c$ meson to charmonium and $D$ mesons in the relativistic quark model,
\href{https://doi.org/10.1103/PhysRevD.68.094020}{Phys. Rev. D \textbf{68}, 094020 (2003)}.

\end{thebibliography}
\end{document}